\documentclass[conference,10pt]{IEEEtran}
\usepackage[margin=1in]{geometry}
\usepackage{amsmath,amssymb,amsfonts}
\usepackage{graphicx}
\usepackage{textcomp}
\usepackage{xcolor}
\usepackage{textgreek}
\usepackage{booktabs}
\usepackage{hyperref}
\usepackage{cleveref}
\usepackage{subcaption}
\usepackage{siunitx}
\usepackage{multirow}
\usepackage{pgfplots}
\pgfplotsset{compat=1.18}
\usepgfplotslibrary{groupplots}

\newcommand{\repo}{\href{https://github.com/arulrhikm/mps-pps-zero-setup-benchmarks}{github.com/arulrhikm/mps-pps-zero-setup-benchmarks}}

\title{Benchmarking Zero-Setup Quantum Circuit Simulators}

\author{Arul Rhik Mazumder\textsuperscript{1,2},  
Hovnatan Karapetyan\textsuperscript{1},  
Mohammed Zuhair Mullath\textsuperscript{1}, \\
Rudy H. Tanin\textsuperscript{1},
Hrant Gharibyan \textsuperscript{1},
Hayk Tepanyan\textsuperscript{1} \\
\\ {\it \textsuperscript{1}BlueQubit Inc,
San Francisco, CA 94105, USA}
\\ {\it \textsuperscript{2}Carnegie Mellon University}
}

\date{}
\begin{document}

\maketitle

\begin{abstract}
Practitioners increasingly rely on hosted simulation environments, but their performance characteristics remain poorly documented. We present a systematic benchmarking study of GPU-accelerated approximate quantum simulation across two widely used methods: matrix product states (MPS) and Pauli path simulation (PPS), comparing BlueQubit (a hosted tool that handles hardware provisioning, simulator configuration, and job orchestration) against AWS Braket, Quantum Rings, Qiskit pauli-prop, and PauliPropagation.jl. For MPS, we find that GPU runtime yields sub-quadratic scaling with bond dimension, with a growing advantage over CPU at increasing scale. For Pauli path simulation on IBM's 127-qubit kicked Ising benchmark, GPUs deliver up to ${\sim}1{,}700\times$ speedup at fine truncation thresholds ($\delta = 2.5 \times 10^{-5}$, 27.6M Pauli terms), and are the only backends that reach accuracy regimes below $\delta = 10^{-5}$, which remained inaccessible to the commodity CPU-based implementations and self-contained SDKs evaluated here. We also provide a reproducible characterization of these simulators across regimes, including tradeoffs that isolated evaluations do not show. All benchmarking code and configurations are in a public GitHub repository.
\end{abstract}


\section{Introduction}
\label{sec:intro}

Classical simulation of quantum circuits is indispensable for algorithm development, hardware validation, and establishing baselines for quantum advantage claims~\cite{preskill2018quantum}. As quantum processors scale to hundreds of qubits (IBM's 127-qubit Eagle, Google's 105-qubit Willow and 53-qubit Sycamore, and Quantinuum's 56-qubit H2 and 98-qubit Helios), the practical limits of exact state-vector simulation ($\mathcal{O}(2^{n})$ for $n$ qubits) force a transition to approximate methods.

In parallel, the ecosystem for running these simulations has shifted. Rather than installing simulation libraries, GPU drivers, and CUDA toolkits on local hardware, practitioners increasingly turn to \emph{zero-setup simulators}: hosted platforms and cloud-accessible backends that allow users to submit circuits and retrieve results without managing any local infrastructure. Services such as BlueQubit, AWS Braket~\cite{aws_braket}, and Quantum Rings~\cite{quantum_rings} follow this model, as do locally installable but self-contained packages like Qiskit pauli-prop (IBM)~\cite{ppsqiskit} and PauliPropagation.jl~\cite{paulipropagationjl}.

Despite the growing reliance on these platforms, no systematic study has compared their end-to-end performance across the two dominant approximate simulation methods, matrix product states (MPS) and Pauli path simulation (PPS), under controlled conditions. Existing benchmarks focus primarily on exact state-vector simulation~\cite{kordzanganeh2023benchmarking, guerreschi2020intel}, and vendor-reported numbers typically cover isolated operations or single circuits rather than systematic sweeps across problem parameters. This gap leaves practitioners without the information needed to select the right backend for a given task.

\textbf{Scope and goals.} This paper provides a standardized, reproducible benchmarking study of zero-setup quantum circuit simulators. We evaluate BlueQubit (CPU and GPU backends for state-vector, MPS, and PPS), AWS Braket SV1, Quantum Rings, Qiskit pauli-prop, and PauliPropagation.jl across a common set of circuit families, parameter sweeps, and statistical protocols. Our primary goals are:
\begin{enumerate}
    \item \textbf{Cross-platform comparison}: Identify which simulators are fastest in which regimes, using the same circuits and metrics throughout.
    \item \textbf{Scaling characterization}: Show how runtime scales with bond dimension ($\chi$), qubit count ($n$), circuit depth ($d$), and truncation threshold ($\delta$), including tradeoffs that are not visible from isolated benchmarks.
    \item A public benchmarking repository\footnote{\repo} containing all code, circuit definitions, and configurations, so that the comparison can be reproduced and extended to new platforms.
\end{enumerate}

\textbf{Key results.} Across all regimes tested, BlueQubit's GPU backend achieves the lowest absolute runtimes. For state-vector simulation, it outperforms AWS Braket SV1 and Quantum Rings by 1--2 orders of magnitude at 30+ qubits. For MPS, we find that the GPU backend exhibits sub-quadratic bond-dimension scaling ($T \propto \chi^{1.49 \pm 0.06}$) compared to near-quadratic CPU scaling ($\chi^{2.03 \pm 0.11}$), a result we attribute to efficient GPU parallelization of tensor contractions and batched SVD operations at large bond dimensions. This produces a growing GPU speedup $S(\chi) \propto \chi^{0.55}$ that favors GPU precisely when simulation is most expensive. However, for low-entanglement circuits (e.g., QFT at $\chi_{\max} = 64$), GPU kernel-launch overhead causes up to $7.5\times$ slowdown relative to CPU, which makes $\chi$, rather than qubit count, the right axis for backend selection. For PPS on IBM's 127-qubit kicked Ising benchmark, the GPU backend delivers up to ${\sim}1{,}700\times$ speedup at fine truncation thresholds and is the only tested backend capable of reaching accuracy regimes beyond the non-monotonic error peak. All benchmarking code, circuit definitions, and configurations are publicly available.\footnote{\repo}

    \section{Background}
    \label{sec:background}

    \subsection{Zero-Setup Quantum Circuit Simulators}
 
    We define a \emph{zero-setup simulator} as any simulation backend that a practitioner can use without installing GPU drivers, CUDA toolkits, or specialized simulation libraries on local hardware. This covers cloud-hosted backends accessed via API (e.g., BlueQubit, AWS Braket SV1) and locally installable, self-contained SDKs that bundle their dependencies (e.g., Quantum Rings, Qiskit pauli-prop, PauliPropagation.jl). In both cases the user's workflow runs from circuit construction to result retrieval with no intermediate infrastructure management. For self-contained SDKs, ``zero-setup'' refers to the software workflow, meaning there are no drivers, toolkits, or specialized libraries to install and configure. It does not refer to the underlying machine, which the user supplies. The host hardware is therefore an explicit experimental variable rather than a fixed property of the backend, and we report it in full (\Cref{sec:setup}) so that software and hardware effects can be read separately.

    \subsection{Matrix Product State Simulation}

    MPS methods decompose an $n$-qubit state $|\psi\rangle$ into a chain of rank-3 tensors $\Gamma^{[k]}$ connected by diagonal matrices $\lambda^{[k]}$ of bond dimension $\chi_{k}$~\cite{vidal2003efficient}:
    \begin{equation}
        |\psi\rangle = \sum \Gamma^{[1]} \lambda^{[1]} \Gamma^{[2]} \lambda^{[2]} \cdots \Gamma^{[n]}
    \end{equation}
    The bond dimension $\chi = \max_k \chi_k$ controls the approximation quality: $\chi = 2^{n/2}$ recovers the exact state, while truncation to $\chi_{\max}$ yields an efficient approximation for states with bounded entanglement entropy.
     
    Applying a two-qubit gate requires: (i) contracting the gate with two adjacent MPS tensors ($\mathcal{O}(\chi^{2})$), and (ii) performing SVD to restore the MPS canonical form ($\mathcal{O}(\chi^{3})$). The SVD step dominates at large $\chi$, yielding an expected runtime per gate of $T_{\text{gate}} \propto \chi^{3}$. However, GPU implementations can alter this balance: tensor contractions map to high arithmetic-intensity GEMM kernels that GPUs execute efficiently, and batched SVD operations reduce per-decomposition overhead. Schieffer et al.~\cite{schieffer2025harnessingcudaqsmpstensor} showed that SVD iterations account for $\sim$70\% of build time on GPU. Following their decomposition, we split MPS execution into a \emph{build phase} (gate-by-gate MPS construction via contraction and SVD) and a \emph{sampling phase} (sequential site measurement to draw bitstrings). The two phases respond differently to parallelization, and we analyze them separately in \Cref{sec:mps-build}.

    \subsection{Pauli Path Simulation}

    PPS methods operate in the Heisenberg picture, evolving an observable $O$ backward through the circuit:
    \begin{equation}
        \langle O \rangle = \text{Tr}[O \cdot \rho] = \text{Tr}[U^\dagger O U \cdot |0\rangle\langle 0|]
    \end{equation}
    The observable is expanded in the Pauli basis, and each gate application potentially doubles the number of Pauli terms. Truncation strategies, which discard terms with coefficients below a threshold $\delta$ or with Pauli weight above a cutoff, keep the computation tractable~\cite{begusic2024fast, aharonov2023polynomial}.
     
    The computational cost scales with the number of retained Pauli paths $N_P$. At threshold $\delta$, $N_{P}(\delta) \propto \delta^{-\alpha}$ for some circuit-dependent exponent $\alpha$. Each Pauli term's update (coefficient truncation, phase accumulation, and rescaling) is independent, which makes PPS naturally data-parallel. GPU implementations exploit this structure via SIMT execution, with efficiency that scales with $N_P$. GPU acceleration primitives for Pauli propagation have recently become available through NVIDIA's cuPauliProp library~\cite{lowell2025cuquantum}, but the end-to-end scaling of speedup with $N_P$ across truncation thresholds has not been systematically characterized in a cross-platform context.

    \subsection{Related Work}

    Kordzanganeh et al.~\cite{kordzanganeh2023benchmarking} benchmarked state-vector simulators including AWS SV1 and QMware, finding QMware fastest below 27 qubits and AWS SV1 advantageous for 28--34 qubits. Chatterjee et al.~\cite{chatterjee2024hamiltonian} used the QED-C framework to benchmark Hamiltonian simulation on CPU/GPU simulators and IBM hardware, identifying classical-quantum crossover points. Kasirajan et al.~\cite{kasirajan2024sycamore} demonstrated that the Quantum Rings SDK simulates Google's 53-qubit Sycamore circuits with XEB fidelity of 0.678 on standard hardware.
 
    On the implementation side, NVIDIA's cuTensorNet~\cite{bayraktar2023cuquantum} provides GPU primitives for MPS simulation, reporting up to $102\times$ speedup for tensor QR at bond dimension 4096 on an A100 GPU versus a 64-core CPU. NVIDIA's cuPauliProp~\cite{lowell2025cuquantum}, released in cuQuantum v25.11, demonstrates orders-of-magnitude speedups on IBM utility circuits. Schieffer et al.~\cite{schieffer2025harnessingcudaqsmpstensor} profiled an MPS implementation on CUDA-Q, finding SVD dominates the build phase; TN-Sim~\cite{hoyt2026tnsim} separately timed contraction vs.\ SVD at fixed $\chi$ and found the same pattern. However, these are micro-benchmarks of individual operations or single-circuit profiling studies, not end-to-end scaling analyses across systematic parameter sweeps, and none compare multiple platforms under a common protocol.
     
    Our work fills the gap between operation-level micro-benchmarks and single-platform evaluations by providing the first standardized, cross-platform benchmarking study of zero-setup quantum simulators, covering both MPS and PPS methods across parameter sweeps in $\chi$, $n$, $d$, and $\delta$.
 
    \section{Experimental Setup}
    \label{sec:setup}

    \subsection{Hardware Specifications}
    All BlueQubit cloud backends run on managed infrastructure with the following hardware:
    \begin{itemize}
    \item \textbf{GPU backends}: NVIDIA A100 GPUs for state-vector and MPS jobs, with backend-specific libraries:
\begin{itemize}
    \item \textbf{SV-GPU}: Accelerated via cuQuantum~\cite{bayraktar2023cuquantum}.
    \item \textbf{MPS-CPU / MPS-GPU}: Built on the Quimb tensor network library~\cite{Gray2018}.
    \item \textbf{PPS-CPU / PPS-GPU}: BlueQubit's proprietary in-house Pauli propagation implementation.
    For the PPS-GPU hardware comparison in \Cref{fig:pps-runtime}, the same BlueQubit PPS-GPU engine was re-run on three accelerators (NVIDIA H100, AMD Instinct MI300X, and NVIDIA H200) so that SKU-to-SKU runtime can be read separately from the cross-framework comparison against Qiskit pauli-prop and PauliPropagation.jl.
\end{itemize}
    \item \textbf{CPU backends}: Intel Xeon Platinum (Sapphire Rapids) processors, with 
backend-specific libraries:
\begin{itemize}
    \item \textbf{SV-CPU}: Based on Google's qsim library~\cite{quantumai2020qsim}.
    \item \textbf{MPS-CPU}: Based on Quimb tensor network library~\cite{Gray2018}.
    \item \textbf{PPS-CPU}: BlueQubit's proprietary in-house coefficient-based Pauli 
    propagation engine.
\end{itemize}
Provisioned in two configurations:
    \begin{itemize}
        \item \texttt{small1}: 2 cores; used for state-vector simulation up to 28 qubits, and for all MPS-CPU and PPS-CPU jobs.
        \item \texttt{small2}: 20 cores; used for state-vector simulation at 29+ qubits.
    \end{itemize}
\end{itemize}
\noindent The dip visible at 29 qubits in the CPU state-vector curves 
(\Cref{fig:bq-qv-sv,fig:cross-platform-d60}) reflects this \texttt{small1}$\to$\texttt{small2} 
hardware transition on the managed platform.

The self-contained SDK baselines (Qiskit pauli-prop and PauliPropagation.jl) were run locally on a Lenovo IdeaPad Slim~7 14IAP7 laptop (model 82SX, x64-based PC) equipped with a 12th Gen Intel Core i7-1260P processor (Alder Lake-P, 2.1\,GHz base, 12 physical cores comprising 4 Performance cores with Hyper-Threading and 8 Efficient cores, for a total of 16 logical processors) and 16.0\,GB of installed physical RAM (15.7\,GB usable), with a Windows-managed page file of 46.0\,GB on the system drive (61.7\,GB total virtual memory). The OS was Windows~11 Home, version 10.0.26200 Build~26200, with BIOS \texttt{LENOVO JHCN39WW} (Dec.\ 2024). No discrete GPU was used; all Qiskit pauli-prop and PauliPropagation.jl runs executed entirely on CPU using each library's default multithreaded settings, without explicit thread affinity or NUMA pinning.

The 16\,GB physical-memory ceiling is the binding constraint for our local PPS-CPU experiments. This is a property of the commodity hardware used here, not of CPU-based PPS in general: large-memory CPU systems, such as workstation- and HPC-class nodes with hundreds of gigabytes to terabytes of RAM, or supercomputers such as Fugaku, Frontier, or Riken-class clusters, should reach substantially finer thresholds without paging, though still subject to the runtime scaling reported in \Cref{fig:pps-runtime}.
    
    \subsection{Simulators Under Test}
     
    We evaluate the following zero-setup simulation backends, covering both cloud-hosted and self-contained SDK-based systems.
     
    \textbf{BlueQubit} (cloud-hosted). Four backends accessed via the BlueQubit cloud API:
    \begin{itemize}
        \item \textbf{SV-CPU}: State-vector on CPU clusters.
        \item \textbf{SV-GPU}: State-vector on NVIDIA GPU clusters via cuQuantum~\cite{bayraktar2023cuquantum}. Multi-GPU scaling beyond 32 qubits.
        \item \textbf{MPS-CPU / MPS-GPU}: Matrix product state on CPU and GPU respectively, with configurable $\chi_{\max}$.
        \item \textbf{PPS-CPU / PPS-GPU}: Pauli path simulation on CPU and GPU with configurable threshold $\delta$.
    \end{itemize}
    
    \textbf{AWS Braket SV1}~\cite{aws_braket} (cloud-hosted). A commercial state-vector simulator supporting up to 34 qubits, accessed via the AWS Braket API.
     
    \textbf{Quantum Rings}~\cite{quantum_rings} (self-contained SDK). An SDK-based simulator supporting 100+ qubits with high cross-entropy benchmarking (XEB) fidelity, installable via pip with no GPU requirements.
     
    \textbf{Qiskit pauli-prop} (IBM)~\cite{ppsqiskit} (self-contained SDK). A Pauli path simulator built on Qiskit~\cite{qiskit}, run locally on CPU.
     
    \textbf{PauliPropagation.jl}~\cite{paulipropagationjl} (self-contained SDK). A Julia-based Pauli propagation library, run locally on CPU.
     
    \noindent\textbf{Scope of cross-platform comparison.}
    Not every backend is evaluated across all three simulation methods. AWS Braket SV1 supports only exact state-vector simulation and provides no MPS or PPS backend via its public API, so it is included in \Cref{sec:sv} only. Braket SV1 benchmarking was also discontinued above 31~qubits, where its runtime exceeded $10^{7}$\,ms and the on-demand cost grew in proportion (\Cref{sec:sv}). Quantum Rings exposes a state-vector interface and does not offer a configurable MPS or PPS mode through its SDK, so it likewise appears only in the state-vector comparison. Both platforms were excluded from the MPS and PPS benchmarks on these technical grounds, not for performance reasons.

    The three results sections therefore cover different backend sets, which we state at the start of each. The state-vector comparison (\Cref{sec:sv}) spans all four state-vector backends: BlueQubit SV-CPU and SV-GPU, AWS Braket SV1, and Quantum Rings. No evaluated zero-setup backend other than BlueQubit exposes a configurable MPS mode, so the MPS benchmarks (\Cref{sec:mps}) compare BlueQubit's CPU and GPU backends only. Because both run the same Quimb-based implementation, this is by construction an \emph{infrastructure} comparison that isolates the effect of hardware on identical code, not a comparison of distinct simulators. The PPS benchmarks (\Cref{sec:pps}) compare BlueQubit PPS-CPU and PPS-GPU against Qiskit pauli-prop and PauliPropagation.jl.

    \textbf{What each comparison measures.} Two kinds of comparison appear in this study, and we label them as such. BlueQubit CPU-versus-GPU results, which cover MPS throughout and PPS where both BlueQubit backends appear, hold the implementation fixed and vary only the hardware, so they isolate the effect of the device. Cross-vendor results, meaning SV1 and Quantum Rings for state-vector and Qiskit pauli-prop and PauliPropagation.jl for PPS, vary implementation and hardware together, so they should be read as end-to-end platform comparisons rather than as measurements of algorithmic efficiency on its own. The provisioning also differs across these backends: BlueQubit's CPU backends run on a bounded core count (\texttt{small1}, 2 cores, for MPS-CPU and PPS-CPU; \texttt{small2}, 20 cores, for large state-vector jobs), and the self-contained SDKs run on the 16\,GB commodity laptop of \Cref{sec:setup}. The CPU-to-GPU speedups we report are relative to these configurations and would be smaller on larger, well-parallelized CPU systems.

    \textbf{Why SV1 as the AWS baseline.} SV1 is AWS's managed, on-demand state-vector simulator, and thus the closest zero-setup analogue to the other state-vector backends compared here: it needs no container, instance selection, or environment setup. AWS also supports GPU-accelerated simulation through Braket Hybrid Jobs~\cite{braket_jobs}, in which the user supplies a container image and selects GPU-backed instances, and it offers other managed simulators such as the density-matrix DM1 and tensor-network TN1 backends. These fall outside our zero-setup scope, since Hybrid Jobs in particular require user-managed containers and instance provisioning, but a GPU-backed AWS workflow could narrow the state-vector gap reported here. We treat this as a limitation of the baseline selection (\Cref{sec:discussion}).

    All backends are evaluated using the same circuit definitions, parameter ranges, and statistical protocol. Where applicable, we distinguish between API-reported runtime (wall-clock time as returned by the platform) and locally measured execution time.
    
    \subsection{Benchmark Circuits}
     
    \begin{enumerate}
        \item \textbf{Quantum Volume (QV).} Random SU(4) circuits generated using Qiskit's \texttt{QuantumVolume} class~\cite{cross2019validating, qiskit2024}, which constructs layers of Haar-random two-qubit unitaries acting on random qubit pairs. These circuits serve as the primary benchmark family across all backends. For state-vector evaluation: $n \in \{16, \ldots, 34\}$ qubits at depths $d \in \{30, 60, 90, 120, 150\}$. For MPS evaluation: extended to $n = 40$ qubits and swept independently over $\chi \in \{32, \ldots, 1536\}$ at fixed $d = 10$, over $d \in \{4, \ldots, 64\}$ at fixed $\chi = 256$, and over $n \in \{16, \ldots, 96\}$ at fixed $d = 10$, $\chi = 256$. The QV circuit at $n = 40$, $d = 10$, $\chi = 256$ is also rerun across shot counts $\texttt{shots} \in \{1, 10, 100, 1000, 5000, 10000\}$ on both CPU and GPU, with build time and total runtime recorded separately. Using one circuit family throughout keeps backends and methods directly comparable.
     
        \item \textbf{Quantum Fourier Transform (QFT).} Full QFT circuits generated using Qiskit's \texttt{QFT} class~\cite{qiskit2024} with approximation degree $k = 0$ (no gate pruning), producing all-to-all controlled-phase gates. We set $\chi_{\max} = 64$ and sweep $n \in \{4, 8, \ldots, 96\}$. QFT generates entanglement that grows only logarithmically with system size, keeping the required bond dimension well below $\chi_{\max}$ throughout. We include QFT as a controlled low-entanglement test case that probes the regime where GPU kernel-launch overhead is expected to dominate compute.
     
        \item \textbf{Kicked Ising (127 qubits).} The transverse-field kicked Ising circuit from IBM's utility experiment~\cite{kim2023evidence} at varying truncation thresholds $\delta$ for PPS benchmarking.
     
        \item \textbf{Peaked Circuits (46 qubits).} A single 46-qubit peaked-circuit \cite{aaronson2024verifiablequantumadvantagepeaked} instance with $1320 \text{ RZZ}$ gates ($\sim$$6646$ total decomposed two-qubit gates). For each $(\text{device}, \chi)$ configuration we sweep MPS bond dimension $\chi \in \{4, 8, 16, 32, 64, 128, 256, 512, 1024, 1536\}$ at $1$ shot, using a \texttt{pauli\_sum} observable consisting of one local $Z_i$ per qubit. The MPS-CPU sweep was terminated at $\chi = 256$: above this point our MPS-CPU runs failed to complete the circuit, so larger $\chi$ values are reported on GPU only. Reconstruction quality is assessed by applying the marginal-attack rule ($\langle Z_i \rangle < 0 \mapsto 1$, else $0$) \cite{gharibyan2025heuristicquantumadvantagepeaked} and comparing the resulting bitstring to the target embedded in the circuit; the overlap percentage serves as a quality proxy for how faithfully the truncated MPS recovers the peaked target distribution at that $\chi$.
    \end{enumerate}
    
    \subsection{Metric and Statistical Protocol}

    For each configuration, we report:
    \begin{itemize}
        \item \textbf{Runtime} ($T$): Wall-clock execution time (or API-reported runtime, when available). Each configuration is executed $N = 5$ times, with only the means plotted in the figures below.
        \item \textbf{Scaling exponent}: Fitted from $T \propto x^{\alpha}$ where $x$ is the scaling variable ($\chi$, $d$, or $n$). We report $\alpha \pm$ standard error with $R^{2}$.
        \item \textbf{Speedup}: $S = T_{\text{baseline}} / T_{\text{fastest}}$ as a function of problem parameters.
        \item \textbf{Phase decomposition}: \texttt{mps\_build\_time} records the MPS construction phase (gate absorption and SVD truncation loop); sampling time is computed as $T_{\text{sampling}} = T_{\text{total}} - T_{\text{build}}$.
        \item \textbf{Fidelity / accuracy}: State fidelity $F = |\langle \psi_{\text{exact}}| \psi_{\text{approx}}\rangle|^{2}$ where tractable; PPS accuracy $\epsilon = |O_{\text{exact}} - O_{\text{PPS}}|$ for observable estimation.
    \end{itemize}
     
    All benchmark code, circuit definitions, and configuration files are publicly available.

\section{Results: State-Vector Baseline}
\label{sec:sv}
 
\begin{figure}[t]
    \centering
    \includegraphics[width=\linewidth]{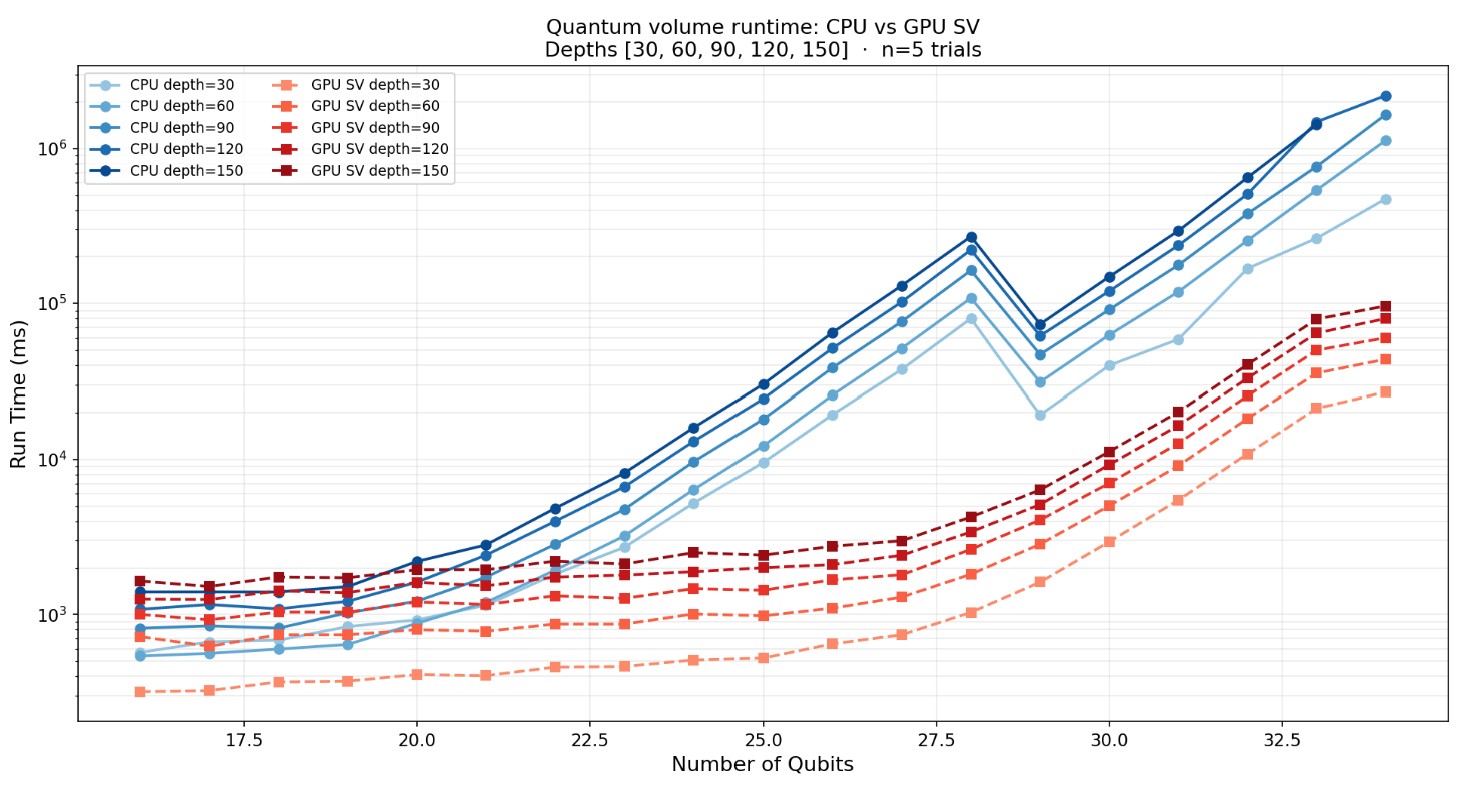}
    \caption{Log-scale runtime (ms) for quantum volume circuits on CPU (solid blue) and GPU (dashed red) across 16--34 qubits and five circuit depths (30--150). GPU outperforms CPU by 1--2 orders of magnitude at higher qubit counts; both scale exponentially. The dip at 29 qubits on CPU curves reflects a hardware transition in the managed platform.}
    \label{fig:bq-qv-sv}
\end{figure}
 
This section evaluates the four state-vector backends: BlueQubit SV-CPU and SV-GPU, AWS Braket SV1, and Quantum Rings. State-vector simulation is the regime where all tested platforms overlap and where practitioners most commonly operate, so we use it to set a baseline. \Cref{fig:bq-qv-sv} shows BlueQubit's CPU and GPU state-vector runtimes across qubit counts and depths.
 
\textbf{GPU advantage at scale.} Across all tested depths ($d = 30$--$150$), GPU runtimes are 1--2 orders of magnitude below CPU at qubit counts $\gtrsim 25$. At smaller counts ($n \lesssim 20$), GPU kernel-launch overhead narrows this gap, with runtimes converging near ${\sim}10^{3}$\,ms.
 
\textbf{Depth sensitivity.} Both CPU and GPU show modest but visible separation across depths at high qubit counts, with deeper circuits incurring proportionally greater cost. This indicates that gate-application time, rather than state-vector allocation, is the dominant cost at $n \leq 34$.
 
\textbf{Exponential wall.} CPU runtimes approach ${\sim}10^{6}$\,ms at 34 qubits; GPU runtimes approach ${\sim}10^{5}$\,ms. Both are prohibitive for iterative workflows, which is what motivates the move to approximate methods beyond ${\sim}30$ qubits.
 
\subsection{Cross-Platform Comparison}

\begin{figure}[t]
    \centering
    \includegraphics[width=\linewidth]{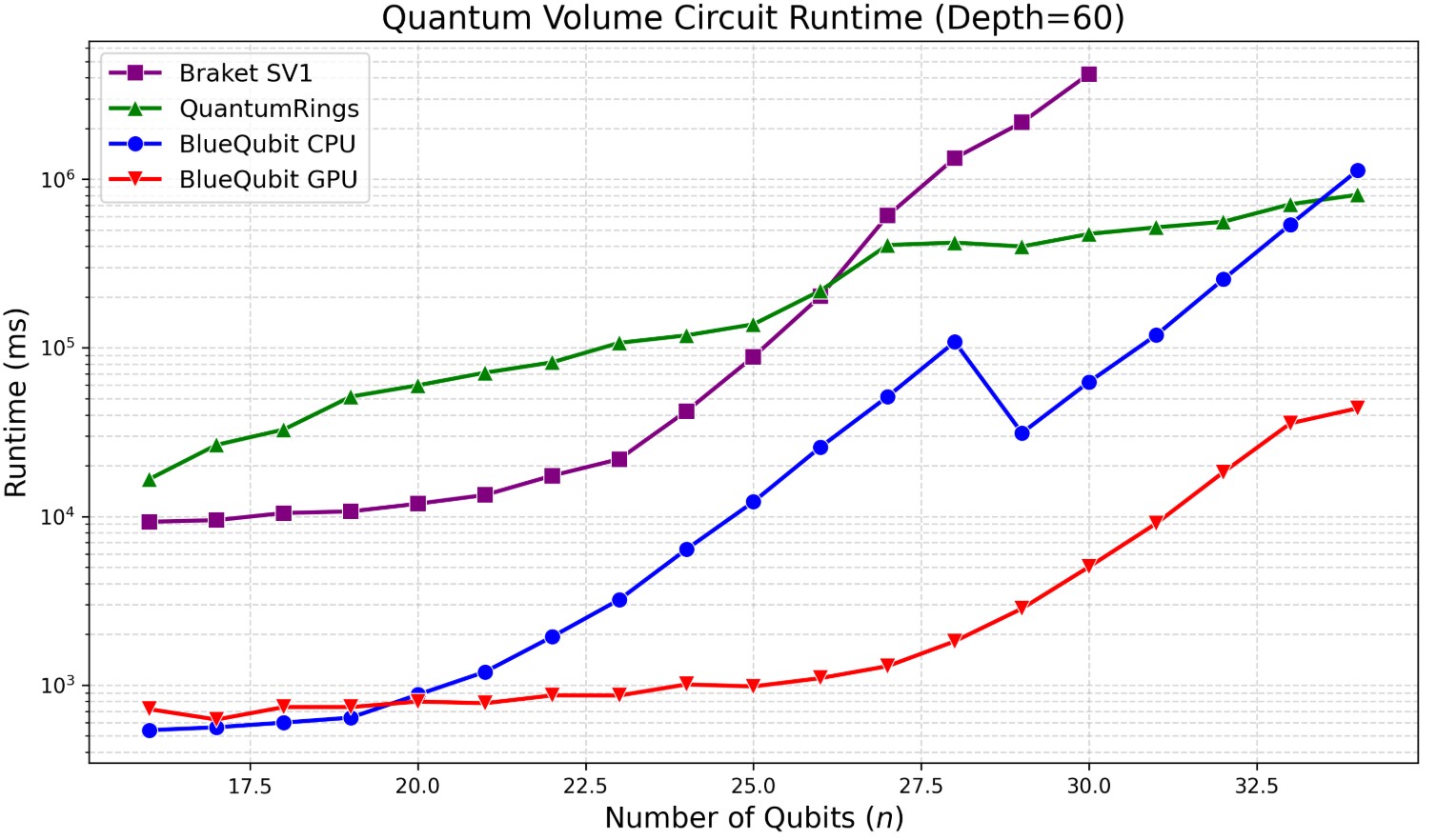}
    \caption{Quantum volume circuit runtime at depth 60 across four backends (16--34 qubits, log scale). BlueQubit GPU (red) is fastest at all qubit counts, reaching ${\sim}4\times10^{4}$\,ms at 34 qubits. BlueQubit CPU (blue) exhibits a characteristic dip near 29 qubits before resuming exponential growth. Braket SV1 (purple) diverges sharply above 27 qubits, surpassing $10^{7}$\,ms above 30 qubits; benchmarking was discontinued beyond 31 qubits due to excessive runtime and cloud cost. Quantum Rings (green) and BlueQubit CPU converge near $10^{6}$\,ms at 34 qubits.}
    \label{fig:cross-platform-d60}
\end{figure}

\Cref{fig:cross-platform-d60} compares all four state-vector backends on the same depth-60 QV circuits. Three findings emerge.
 
First, BlueQubit GPU is the fastest state-vector backend at all tested qubit counts, outperforming Braket SV1 and Quantum Rings by 1--2 orders of magnitude at 30+ qubits and BlueQubit CPU by roughly $2\times$ at 34 qubits.
 
Second, platform differences grow with problem size. At 20 qubits, all backends complete within a few seconds; at 34 qubits, the spread exceeds three orders of magnitude. This is why benchmarking at scale matters more than small-circuit comparisons.
 
Third, Braket SV1 exhibits steep cost scaling above 27 qubits, with runtimes surpassing $10^{7}$\,ms at 30+ qubits. Benchmarking was discontinued beyond 31 qubits due to both runtime and cloud cost considerations.
 
The QV circuit family is carried forward into the MPS benchmarks of \Cref{sec:mps}, with $\chi$ added as a sweep axis.

\section{Results: MPS Benchmarks}\label{sec:mps}

The MPS benchmarks compare BlueQubit's CPU and GPU backends only. As noted in \Cref{sec:setup}, no other evaluated zero-setup backend exposes a configurable MPS mode, so this section is a same-implementation, CPU-versus-GPU infrastructure comparison rather than a comparison across simulators. All MPS benchmarks use quantum volume circuits (random SU(4) layers) unless otherwise noted, with $N = 5$ independent runs per configuration. Using the same circuit family as the state-vector baseline (\Cref{sec:sv}) ensures that observed differences reflect simulator behavior rather than circuit-specific structure.

\subsection{Bond Dimension Scaling}
\label{sec:mps-chi}
 
Bond dimension $\chi$ controls MPS approximation quality and is the primary driver of computational cost. GPU parallelization of tensor contractions and batched SVD operations puts runtime in a qualitatively different scaling regime than CPU execution.

\begin{figure}[t]
    \centering
    \includegraphics[width=\linewidth]{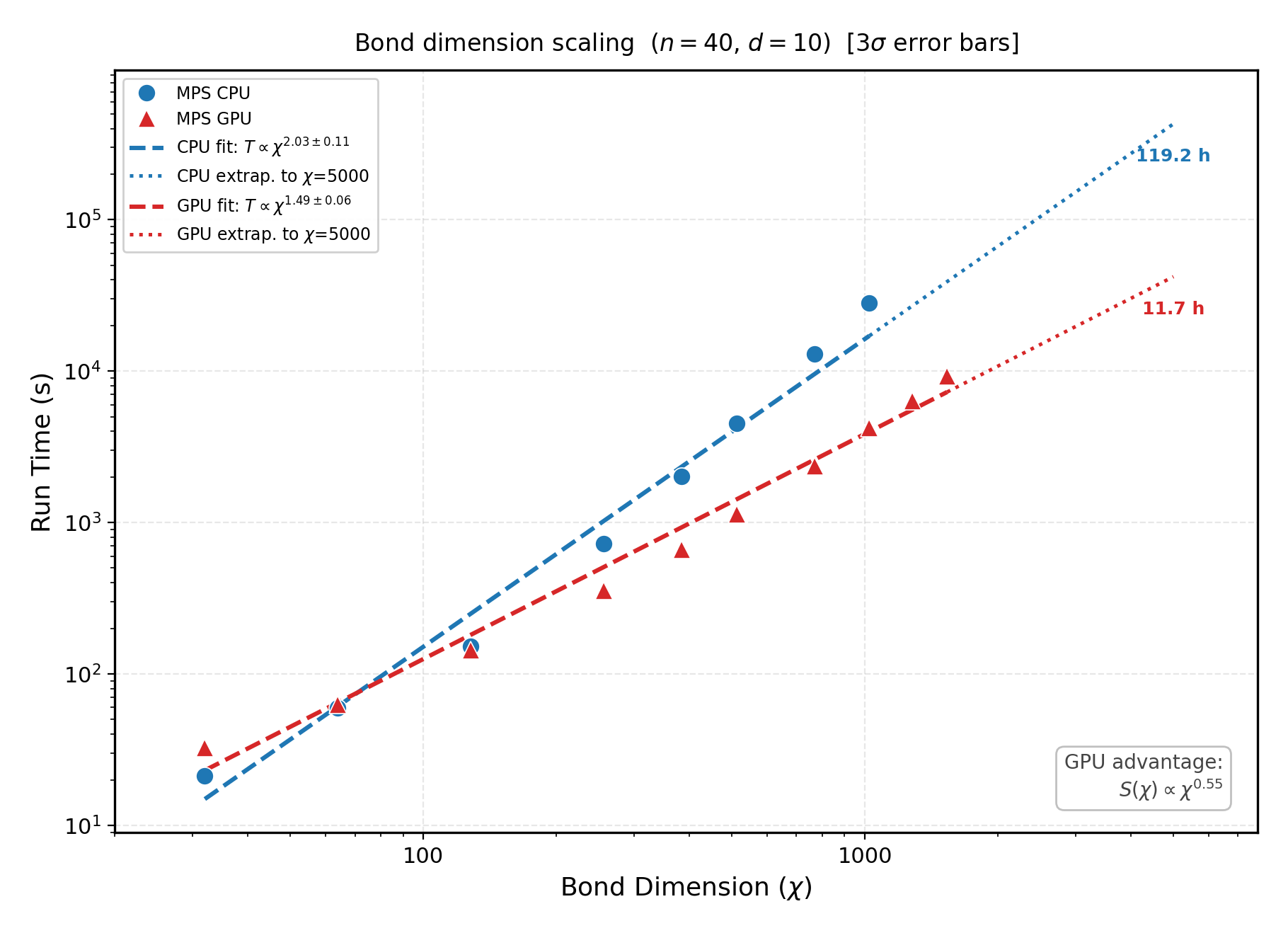}
    \caption{MPS runtime vs.\ bond dimension at $n = 40$ qubits, $d = 10$ layers (log--log scale). CPU scaling ($T \propto \chi^{2.03 \pm 0.11}$) approaches the theoretical $\chi^{2}$ for gate-contraction-dominated regimes. GPU scaling ($T \propto \chi^{1.49 \pm 0.06}$) is sub-quadratic, reflecting efficient parallelization of contraction and SVD operations at large tensor sizes. The GPU advantage grows with $\chi$ as $S(\chi) \propto \chi^{0.55}$. Dotted lines extrapolate both fits to $\chi = 5{,}000$, projecting runtimes of 119.2\,h (CPU) vs.\ 11.7\,h (GPU).}
    \label{fig:mps-chi}
\end{figure}

\Cref{fig:mps-chi} presents the central MPS result. At $n = 40$ qubits and depth 10, we measure:
\begin{equation}
    T_{\text{CPU}}(\chi) = a_{C} \cdot \chi^{2.03 \pm 0.11}
\end{equation}
\begin{equation}
    T_{\text{GPU}}(\chi) = a_{G} \cdot \chi^{1.49 \pm 0.06}
\end{equation}

\textbf{CPU exponent.} The CPU exponent of $2.03$ is consistent with a contraction-dominated regime where the $\mathcal{O}(\chi^2)$ gate absorption cost exceeds the $\mathcal{O}(\chi^3)$ SVD truncation cost at the bond dimensions tested ($\chi \leq 1536$); the transition to SVD-dominated $\chi^3$ scaling is expected at larger $\chi$ or greater depth~\cite{schollwock2011density}.

\textbf{GPU exponent.} The GPU exponent of $1.49$ is sub-quadratic, below both the CPU value and the naive $\chi^{2}$ expectation. This reduction is attributable to two well-understood sources of GPU parallelism. First, the tensor contractions that dominate at these bond dimensions map to GEMM operations, where GPU throughput scales favorably with matrix size. Second, batched SVD execution reduces per-decomposition overhead as tensor dimensions grow. Published profiling of comparable MPS implementations confirms that SVD accounts for $\sim$70\% of GPU build time~\cite{schieffer2025harnessingcudaqsmpstensor}, and NVIDIA's cuTensorNet reports order-of-magnitude speedups for both tensor SVD and gate-split primitives at large bond dimension~\cite{bayraktar2023cuquantum}. The net effect is that the GPU's relative efficiency increases with $\chi$, producing a lower effective scaling exponent.

\textbf{Growing advantage.} Because $1.49 < 2.03$, the GPU speedup
\begin{equation}
    S(\chi) = \frac{T_\text{CPU}}{T_{\text{GPU}}} \propto \chi^{0.55}
\end{equation}
increases with bond dimension: the GPU advantage is largest precisely when simulation is most expensive. Extrapolating the power-law fits to $\chi = 5000$ projects $\sim$11.7\,h on GPU vs.\ $\sim$119.2\,h on CPU (a $\sim$$10\times$ projected speedup), though we caution that this is illustrative only: at sufficiently large $\chi$ the $\mathcal{O}(\chi^3)$ SVD cost must eventually dominate on both backends, and the effective exponent should rise. Our data does not yet show this transition, suggesting it occurs beyond $\chi = 1536$ for the circuit depths tested.
 
\subsection{Build vs.\ Sampling Phase Decomposition}
\label{sec:mps-build}
 
\begin{figure}[t]
    \centering
    \includegraphics[width=\linewidth]{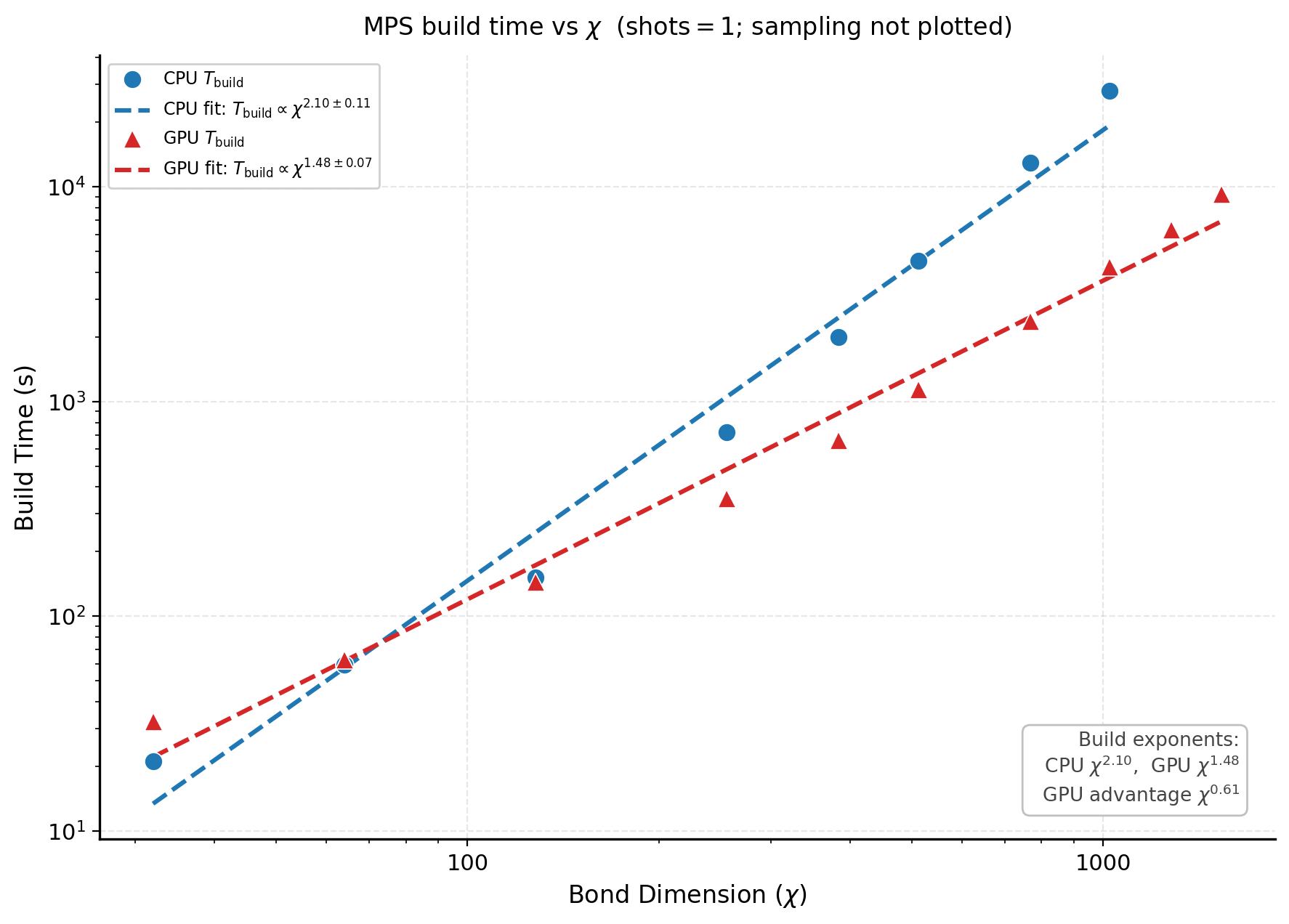}
    \caption{MPS build-phase runtime vs.\ bond dimension at $\texttt{shots} = 1$, isolating the MPS construction phase from sampling overhead. CPU build time scales as $\chi^{2.10 \pm 0.11}$; GPU build time scales as $\chi^{1.48 \pm 0.07}$, yielding a growing GPU advantage of $\chi^{0.61}$. Sampling time is negligible at \texttt{shots}$=1$ and is not plotted.}
    \label{fig:mps-build}
\end{figure}

To understand where the sub-quadratic GPU scaling originates, we decompose the total MPS runtime into its constituent phases. At \texttt{shots}$=1$, sampling time is negligible compared to build time at all tested bond dimensions, confirming that the end-to-end scaling reported above reflects the build phase exclusively.

\textbf{Build phase exponents.} The build-phase scaling exponents,
\begin{equation}
    T_{\text{build}}^{\text{CPU}} \propto \chi^{2.10 \pm 0.11}, \quad
    T_{\text{build}}^{\text{GPU}} \propto \chi^{1.48 \pm 0.07},
\end{equation}
match the end-to-end exponents ($2.03$ and $1.49$, respectively) to within measurement uncertainty, confirming that the scaling behavior is localized entirely to the build phase: the gate-by-gate tensor contraction and SVD loop. The GPU advantage in the build phase scales as $\chi^{0.61}$, consistent with the end-to-end figure of $\chi^{0.55}$. The GPU advantage thus comes from faster circuit construction, not from faster sampling: for many-shot workflows the build cost is amortized; for many-circuit workflows the build-phase speedup translates directly to end-to-end improvement.
 
\begin{figure}[t]
    \centering
    \includegraphics[width=\linewidth]{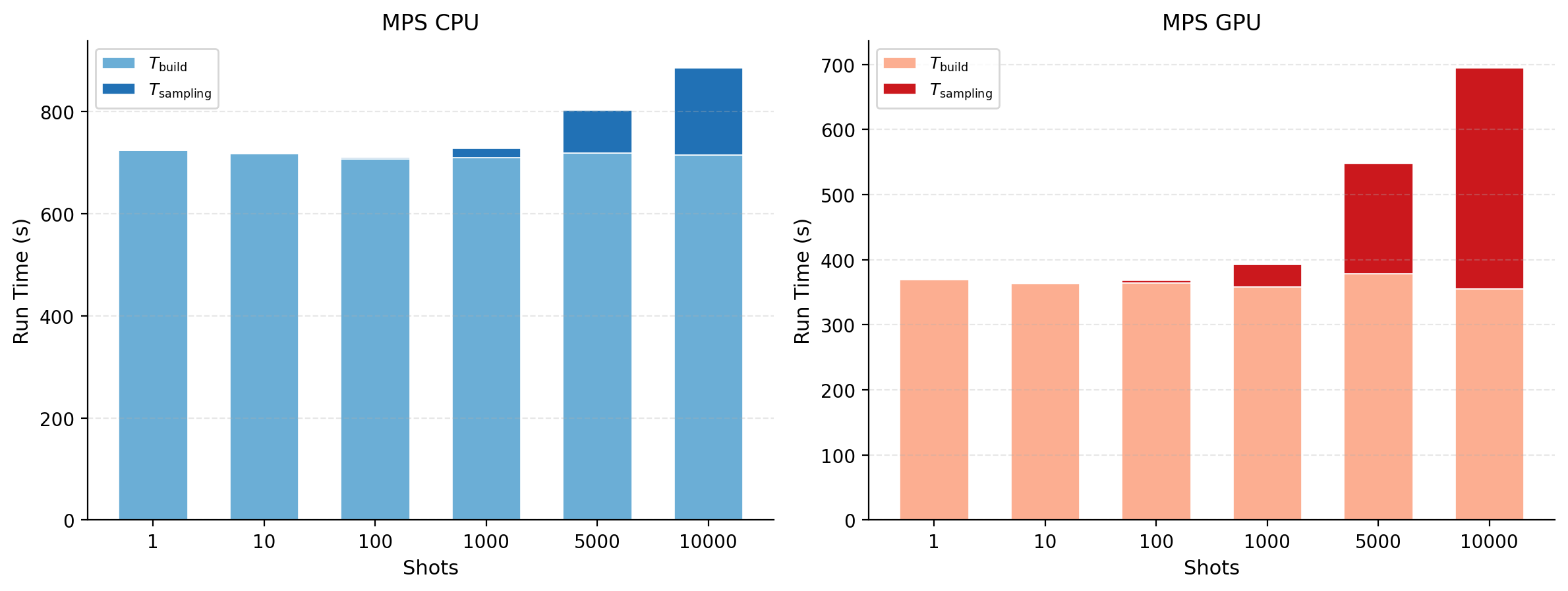}
    \caption{MPS build vs.\ sampling time as a function of shot count at fixed $n = 40$, $d = 10$, $\chi = 256$, on CPU (left) and GPU (right). Build time (light) is shot-independent on both backends (${\sim}715$\,s on CPU and ${\sim}360$\,s on GPU), confirming that circuit construction and sampling are separable phases. Sampling time (dark) grows with shot count substantially at $\geq 5000$ shots, particularly on GPU where sampling rises more steeply than on CPU.}
    \label{fig:mps-shots}
\end{figure}
 
\textbf{Shots scaling.} \Cref{fig:mps-shots} decomposes runtime across $\texttt{shots} \in \{1, 10, 100, 1000, 5000, 10000\}$ at fixed $n = 40$, $d = 10$, $\chi = 256$. Build time is shot-independent on both backends (${\sim}715$\,s CPU, ${\sim}360$\,s GPU), confirming structural separability of construction and sampling. Sampling time becomes a substantial cost above ${\sim}5000$ shots, growing more steeply on GPU than CPU; consequently, the end-to-end GPU speedup erodes from $2.0\times$ at $\texttt{shots} = 1$ to ${\sim}1.3\times$ at $\texttt{shots} = 10000$. GPU is the right choice for construction-bound workflows; the gap narrows for sampling-bound workloads.

\subsection{Qubit Scaling}
\label{sec:mps-qubit}

\begin{figure}[t]
    \centering
    \includegraphics[width=\linewidth]{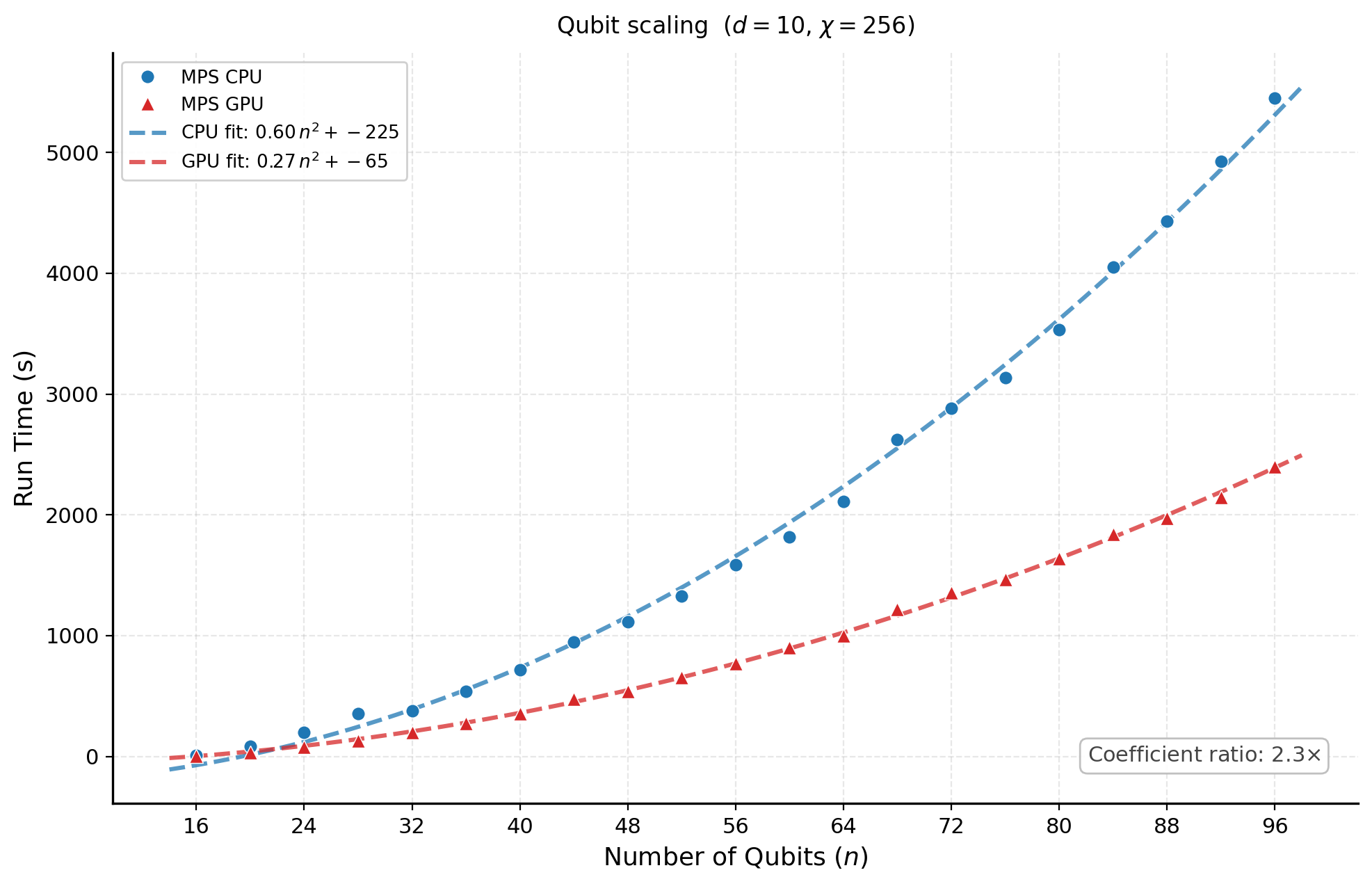}
    \caption{MPS runtime vs.\ qubit count at fixed $d = 10$, $\chi = 256$, across $n \in \{16, \ldots, 96\}$. Both CPU and GPU scale quadratically with $n$ as expected (more qubits $=$ more gates at constant depth). CPU fit: $0.60\,n^{2} - 225$ (s); GPU fit: $0.27\,n^{2} - 65$ (s). The GPU quadratic coefficient is $2.2\times$ smaller, giving a consistent speedup across qubit counts.}
    \label{fig:mps-qubit}
\end{figure}

\Cref{fig:mps-qubit} shows runtime versus qubit count at fixed depth $d=10$ and bond dimension $\chi = 256$. Both backends follow quadratic growth, consistent with the $\mathcal{O}(n \cdot d \cdot \chi^{2})$ gate cost where $n \cdot d$ counts the total number of two-qubit gates at fixed depth.
\begin{equation}
    T_{\text{CPU}}(n) = 0.60\,n^{2} - 225~\text{s}, \quad
    T_{\text{GPU}}(n) = 0.27\,n^{2} - 65~\text{s}.
\end{equation}
The ratio of quadratic coefficients, $0.60 / 0.27 = 2.2\times$, gives the qubit-count-independent GPU speedup at this $\chi$. The negative intercepts reflect a threshold regime at small $n$ where circuit size is insufficient to saturate GPU parallelism. Both fits achieve $R^{2} > 0.99$ across the full range $n \in \{16, \ldots, 96\}$.

\subsection{Circuit Family Dependence: QFT as a Low-Entanglement Control}
 
\begin{figure}[t]
    \centering
    \includegraphics[width=\linewidth]{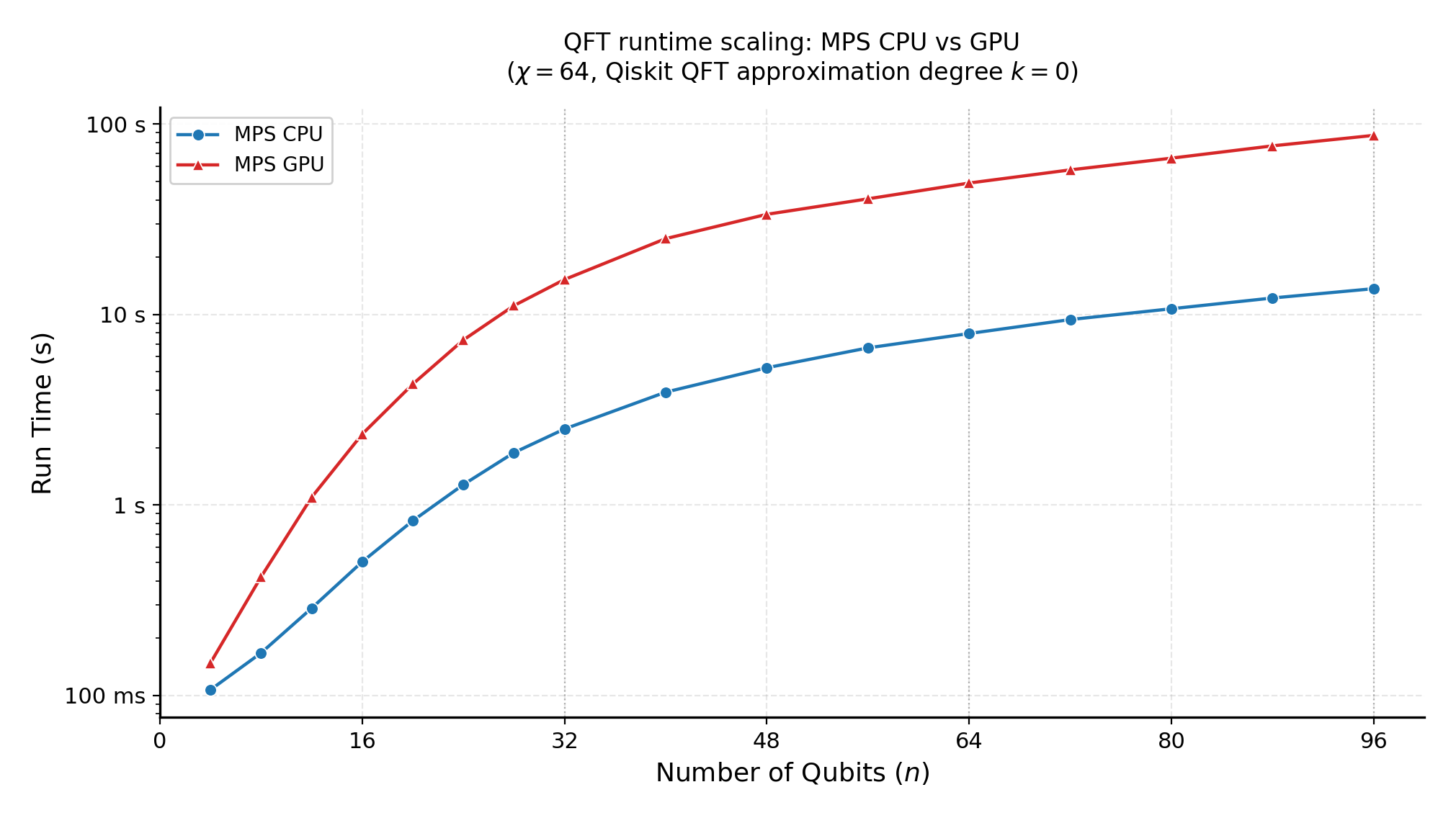}
    \caption{MPS runtime vs.\ qubit count for full QFT circuits ($k = 0$, $\chi_{\max} = 64$), $n \in \{4, \ldots, 96\}$. Unlike the QV results, GPU (red) is slower than CPU (blue) across the entire qubit range, with the gap widening monotonically from near-parity at $n = 4$ to a ${\sim}7.5\times$ GPU slowdown at $n = 96$ (${\sim}90$\,s GPU vs.\ ${\sim}12$\,s CPU). Both backends exhibit sub-linear growth in $n$, consistent with the logarithmic entanglement structure of QFT.}
    \label{fig:mps-qft}
\end{figure}

A natural question is whether the GPU advantage observed on quantum volume circuits generalizes to all circuit families. To test this, we evaluate full QFT circuits (Qiskit approximation degree $k=0$) up to $n = 96$ qubits at $\chi_{\max} = 64$. The result, shown in \Cref{fig:mps-qft}, is the opposite of the QV finding: GPU is slower than CPU across the entire range, by up to $7.5\times$ at $n = 96$.

\textbf{Why QFT favors CPU.} The QFT's controlled-phase gate structure generates entanglement that grows only as $\mathcal{O}(\log n)$~\cite{vidal2003efficient}, so the MPS bond dimension stays well below $\chi_{\max} = 64$ throughout the circuit. At these small tensor sizes, each GPU kernel launch, device-side synchronization, and memory transfer incurs a fixed overhead that exceeds the useful computation per gate. Across the $\mathcal{O}(n^2)$ gates in a QFT circuit, this overhead accumulates, which is what produces the monotonically widening gap in \Cref{fig:mps-qft}.

\textbf{Consistency with the scaling model.} The $S(\chi) \propto \chi^{0.55}$ model from \Cref{sec:mps-chi} was derived in the regime where GPU compute exceeds overhead. Extrapolating downward to $\chi = 64$ predicts a GPU advantage of only ${\sim}10\%$, which is clearly not realized. The observed $7.5\times$ slowdown confirms that $\chi = 64$ lies in the overhead-dominated regime where the power-law model breaks down. Taken together, the QV and QFT results bound the break-even point: GPU overhead exceeds GPU compute benefit at $\chi = 64$ (slowdown), while GPU is clearly faster at $\chi \approx 128$--$256$ (speedup of $2\text{--}3\times$ from \Cref{fig:mps-chi}).

\textbf{Practical guidance.} These results define a concrete backend selection criterion based on circuit entanglement structure rather than qubit count:
\begin{itemize}
    \item \textit{Low-entanglement circuits} ($\chi \lesssim 128$, e.g.\ QFT, shallow circuits, product-state preparations): use CPU. GPU kernel-launch overhead dominates at small tensor sizes.
    \item \textit{High-entanglement circuits} ($\chi \gtrsim 256$, e.g.\ random QV circuits at moderate depth, kicked Ising): use GPU. The speedup exceeds $3\times$ at $\chi = 256$ and grows as $\chi^{0.55}$.
\end{itemize}
This rule is independent of qubit count: the QFT slowdown persists to $n = 96$ qubits, so $\chi$ and not $n$ is the axis that should drive backend selection.

\subsection{Depth Scaling}
\label{sec:mps-depth}

\begin{figure}[t]
    \centering
    \includegraphics[width=\linewidth]{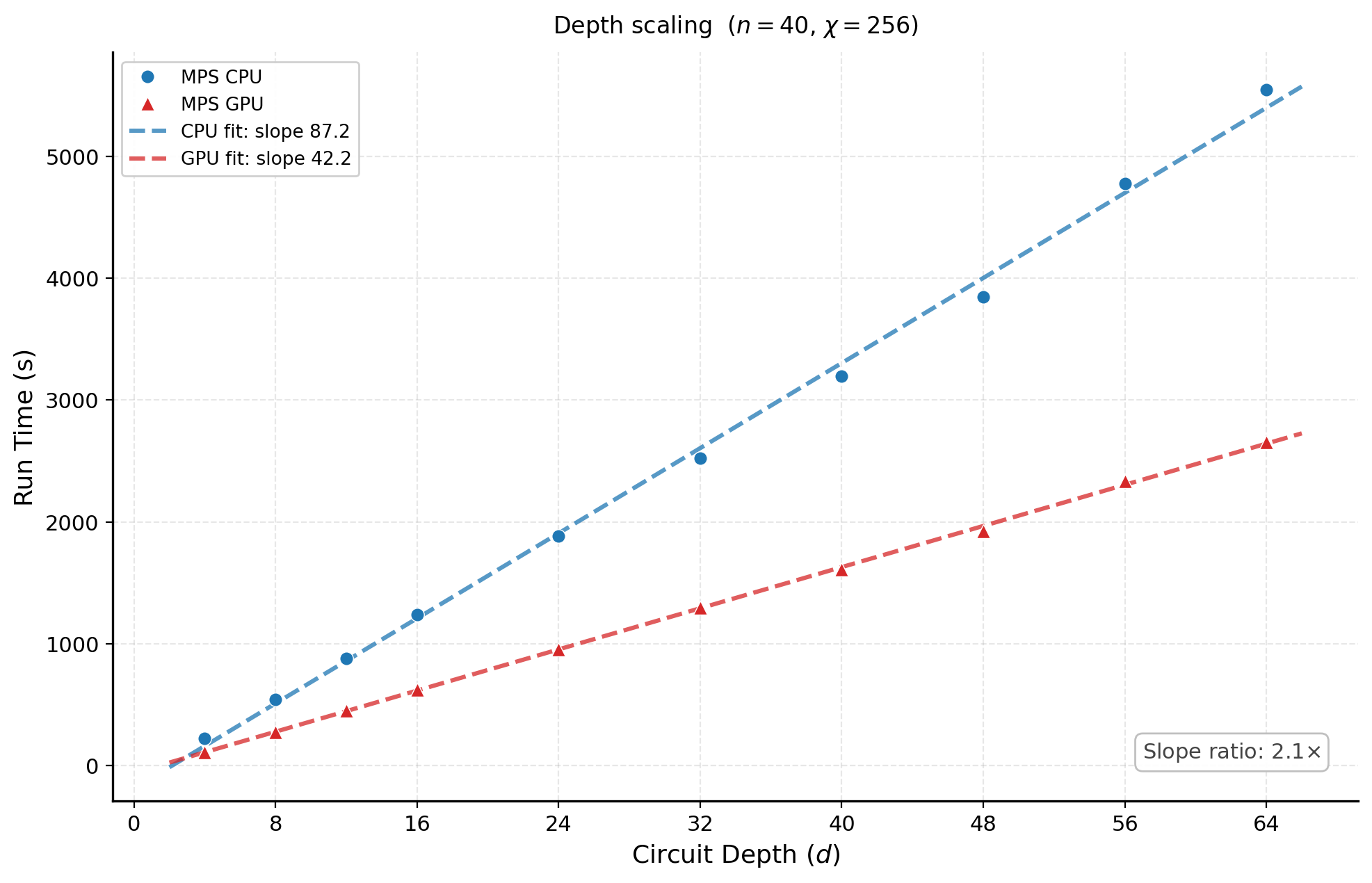}
    \caption{MPS runtime vs.\ circuit depth at $n = 40$, $\chi = 256$, across depths $d \in \{4, \ldots, 64\}$. Both backends scale linearly with depth, consistent with the $\mathcal{O}(d)$ gate count at fixed $n$. CPU slope: $87.2$\,s/layer; GPU slope: $42.2$\,s/layer. The slope ratio ($2.1\times$) is constant across the full depth range, indicating that the per-gate speedup is depth-independent.}
    \label{fig:mps-depth}
\end{figure}

\Cref{fig:mps-depth} shows runtime versus circuit depth at $n = 40$, $\chi = 256$. Both CPU and GPU scale linearly with depth, as expected from the $\mathcal{O}(d)$ gate count at fixed $n$ and $\chi$:
\begin{equation}
    T_{\text{CPU}}(d) = 87.2\,d + b_{C}, \quad
    T_{\text{GPU}}(d) = 42.2\,d + b_{G}.
\end{equation}
The slope ratio of $87.2 / 42.2 = 2.1\times$ is constant across the tested depth range, confirming that the GPU advantage is a per-gate effect that does not depend on circuit depth. This is consistent with the qubit-scaling result ($2.2\times$ at $\chi = 256$) and indicates that the $\sim$2$\times$ speedup at this bond dimension is a stable baseline. The full benefit of GPU parallelization emerges at higher bond dimensions, where the sub-quadratic exponent advantage dominates (\Cref{sec:mps-chi}).

\subsection{Runtime Scaling on Peaked Circuits: CPU vs.\ GPU}
\label{sec:mps-peaked}
\begin{figure}[t]
    \centering
    \includegraphics[width=\linewidth]{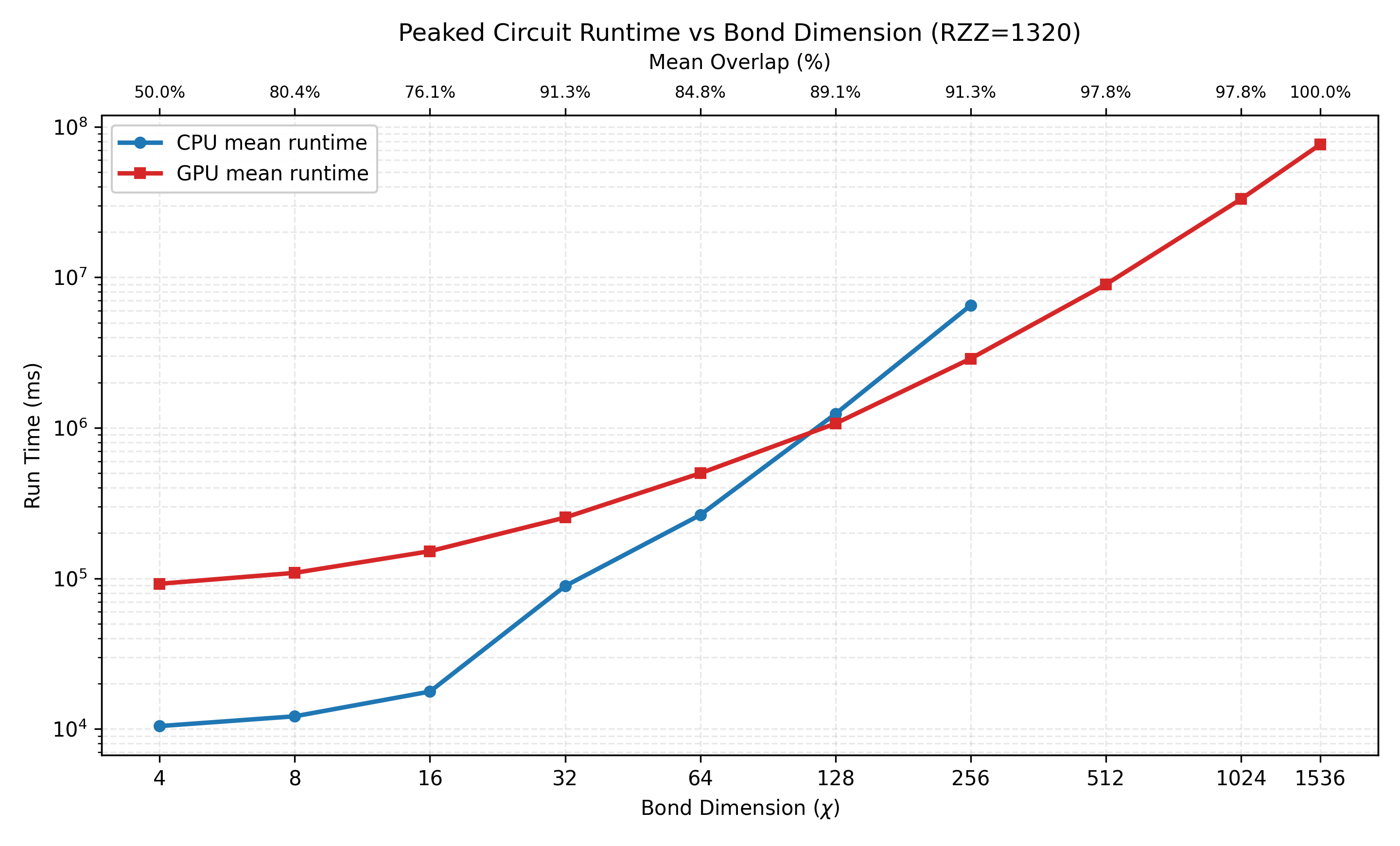}
    \caption{MPS runtime vs.\ bond dimension $\chi$ on a 46-qubit peaked circuit with $1320$ $\mathrm{RZZ}$ gates (decomposed into ${\sim}6646$ two-qubit gates), comparing CPU (blue) and GPU (red) backends. The CPU sweep terminates at $\chi = 256$: above this point our MPS-CPU runs failed to complete the circuit, so larger $\chi$ values are GPU-only. Top-axis labels report the mean overlap (\%) between the marginal-attack reconstructed bitstring and the embedded target at each $\chi$.}
    \label{fig:mps-peaked-runtime}
\end{figure}

Figure~\ref{fig:mps-peaked-runtime} benchmarks MPS simulation on the 46-qubit peaked-circuit instance described in Section~\ref{sec:setup} ($\mathrm{RZZ} = 1320$), sweeping bond dimension $\chi$ from $4$ to $1536$ on GPU and from $4$ to $256$ on CPU.

Two regimes are visible. At low bond dimension ($\chi \leq 16$), the CPU is roughly an order of magnitude faster than the GPU (around $10^4$\,ms versus $10^5$\,ms) because the per-step linear-algebra work is too small to amortize GPU kernel-launch and host-device transfer overhead. The GPU curve is nearly flat in this regime, indicating that runtime is dominated by fixed overhead rather than tensor contraction cost. As $\chi$ grows, contraction and SVD cost (which scales as $\mathcal{O}(\chi^3)$ per bond) begins to dominate, and the CPU curve steepens sharply: between $\chi = 16$ and $\chi = 256$, CPU runtime grows by roughly three orders of magnitude, while GPU runtime grows by only about $1.5$ orders. The crossover occurs near $\chi \approx 128$, beyond which the GPU is increasingly favored. At $\chi = 256$ the CPU is already $\sim$$2\times$ slower than the GPU, and is also where the CPU sweep terminates (see \Cref{sec:setup}). The GPU continues to $\chi = 1536$, where it requires ${\sim}8 \times 10^{7}$\,ms (${\sim}22$\,h) per run.

The top-axis overlap values trend upward with $\chi$, from $50.0\%$ at $\chi = 4$ (consistent with marginal-attack reconstruction at chance under severe truncation) to $100.0\%$ at $\chi = 1536$. The trajectory is not strictly monotonic at intermediate $\chi$: increasing $\chi$ refines some local marginals while flipping others through truncation-induced sign changes, so single-instance overlap fluctuates by a few percent before settling. The overall trend confirms that larger bond dimension yields a better representation of the peaked target distribution, and that reaching near-perfect overlap on this 46-qubit instance requires bond dimensions where only the GPU backend remains practical.
 
\section{Results: PPS Benchmarks}
\label{sec:pps}
This section benchmarks PPS implementations on the 127-qubit kicked Ising model from IBM's quantum utility experiment~\cite{kim2023evidence}, measuring $\langle Z_{62}\rangle$ at $n_{\text{Trotter}} = 20$ Trotter steps with $\theta_h = \pi/4$. We compare BlueQubit PPS-CPU against two CPU-only self-contained SDKs, Qiskit pauli-prop and PauliPropagation.jl (both on the 16\,GB commodity laptop of \Cref{sec:setup}), and BlueQubit PPS-GPU on three accelerators (NVIDIA H100, AMD Instinct MI300X, and NVIDIA H200) running the same in-house engine. We additionally overlay NVIDIA's GPU-accelerated cuPauliProp~\cite{lowell2025cuquantum} at the three thresholds for which speedups over Qiskit pauli-prop have been published. The comparison spans truncation thresholds over four orders of magnitude ($\delta \in [2.89 \times 10^{-6}, 10^{-2}]$), corresponding to Pauli-term counts from 664 to ${\sim}1.5$B.
 
\begin{figure}[t]
    \centering
    \includegraphics[width=\linewidth]{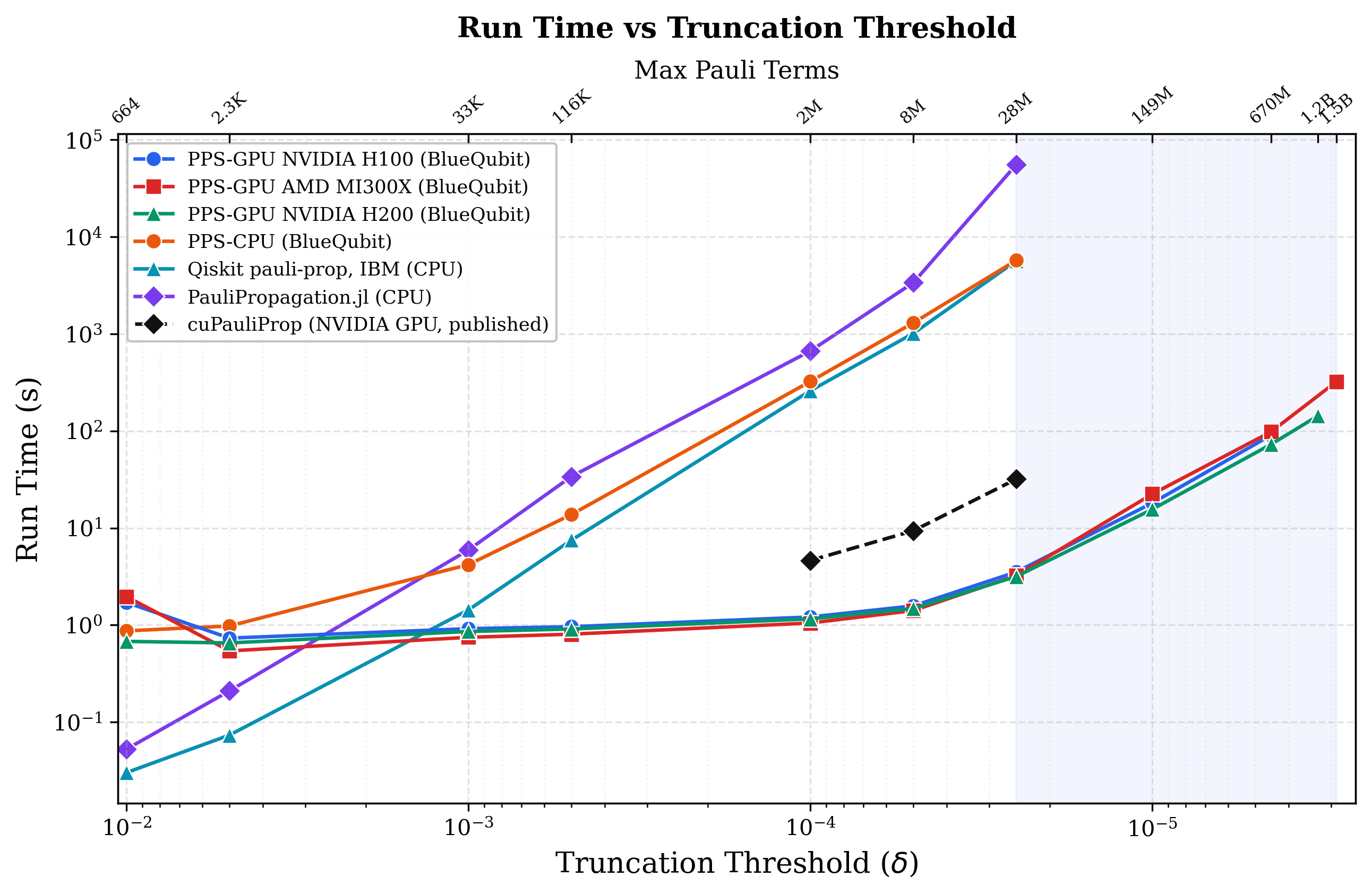}
    \caption{PPS runtime vs.\ truncation threshold $\delta$ on the 127-qubit kicked Ising circuit. BlueQubit PPS-GPU is shown separately on NVIDIA H100, AMD MI300X, and NVIDIA H200; the CPU baselines are BlueQubit PPS-CPU together with the two CPU-only SDKs, Qiskit pauli-prop and PauliPropagation.jl. The dashed black series is NVIDIA's GPU-accelerated cuPauliProp; its runtimes are not measured here, but inferred from the speedups over Qiskit pauli-prop reported in~\cite{lowell2025cuquantum}, which is why it is drawn dashed and spans only three thresholds. The secondary top axis shows the corresponding maximum Pauli-term count. The shaded region marks thresholds reachable only by GPU in this study. At coarse thresholds ($\delta \geq 5 \times 10^{-3}$), CPU implementations are faster due to GPU kernel-launch overhead; the crossover occurs near $\delta \approx 10^{-3}$. At the finest threshold reachable by all CPU backends ($\delta = 2.5 \times 10^{-5}$, 27.6M~Paulis), all three GPU SKUs complete in ${\sim}3$--$3.5$\,s (${\sim}1{,}600\times$--$1{,}800\times$ vs.\ BlueQubit CPU). Beyond $\delta = 10^{-5}$, only GPU continues: H100 to $\delta = 4.5\times10^{-6}$ (${\sim}670$M Paulis), H200 to $\delta = 3.28\times10^{-6}$ (${\sim}1.21$B), and MI300X to $\delta = 2.89\times10^{-6}$ (${\sim}1.53$B).}
    \label{fig:pps-runtime}
\end{figure}
 
\begin{figure}[t]
    \centering
    \includegraphics[width=\linewidth]{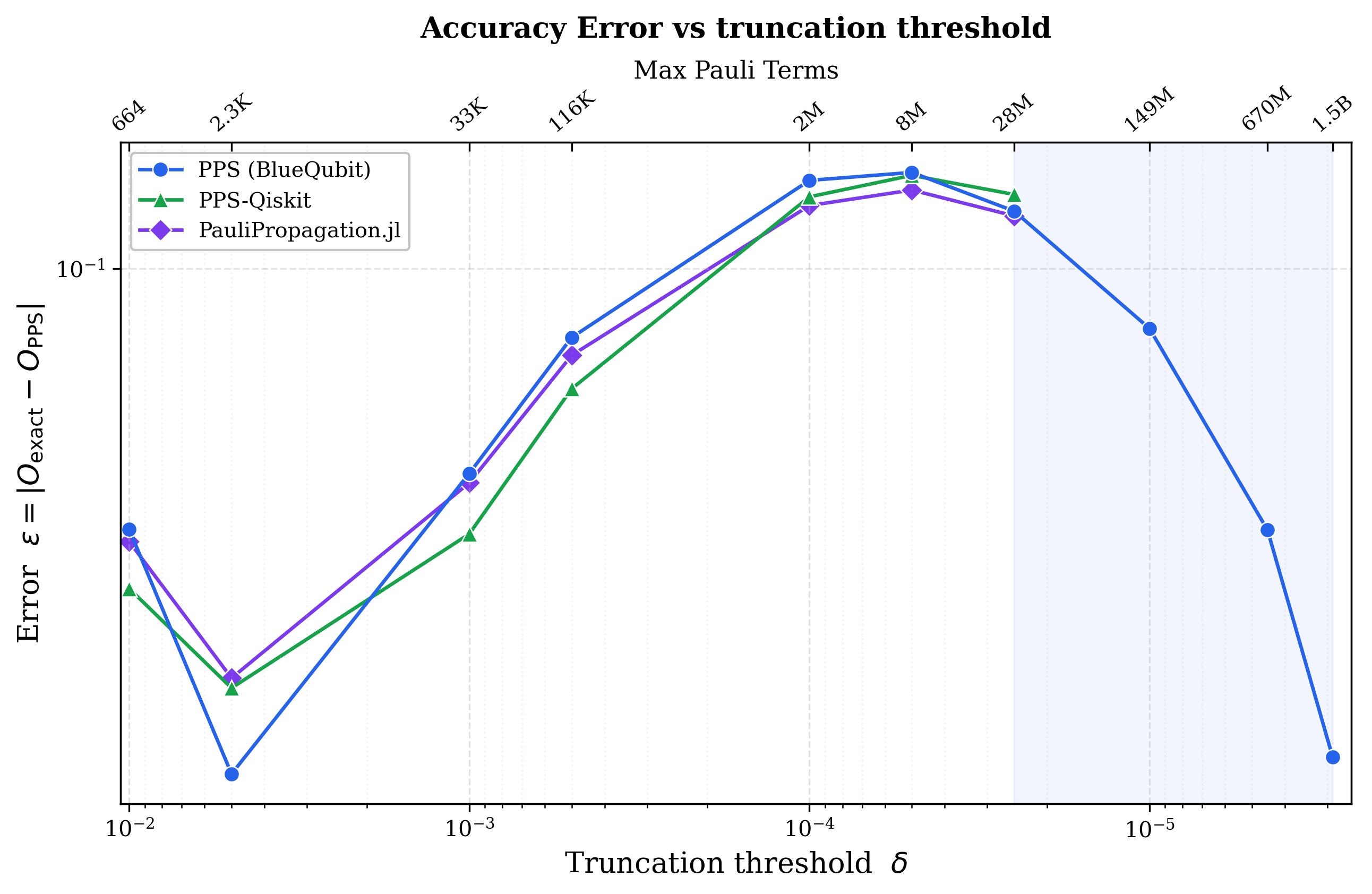}
    \caption{Accuracy error $\epsilon = |O_{\mathrm{exact}} - O_{\mathrm{PPS}}|$ vs.\ truncation threshold, with $O_{\mathrm{exact}} = 0.2955$ from converged tensor network simulations~\cite{begusic2024converged, tindall2024efficient}. The error is non-monotonic: it dips to $\epsilon \approx 0.015$ at $\delta = 5\times10^{-3}$, then rises to a peak of $\epsilon \approx 0.14$ near $\delta = 5\times10^{-5}$, then decreases through the GPU-only regime to $\epsilon \approx 0.016$ at the finest threshold tested. The accuracy recovery occurs entirely in the shaded region that only the GPU backend can reach in our evaluation.}
    \label{fig:pps-accuracy}
\end{figure}
 
\begin{figure}[t]
    \centering
    \includegraphics[width=\linewidth]{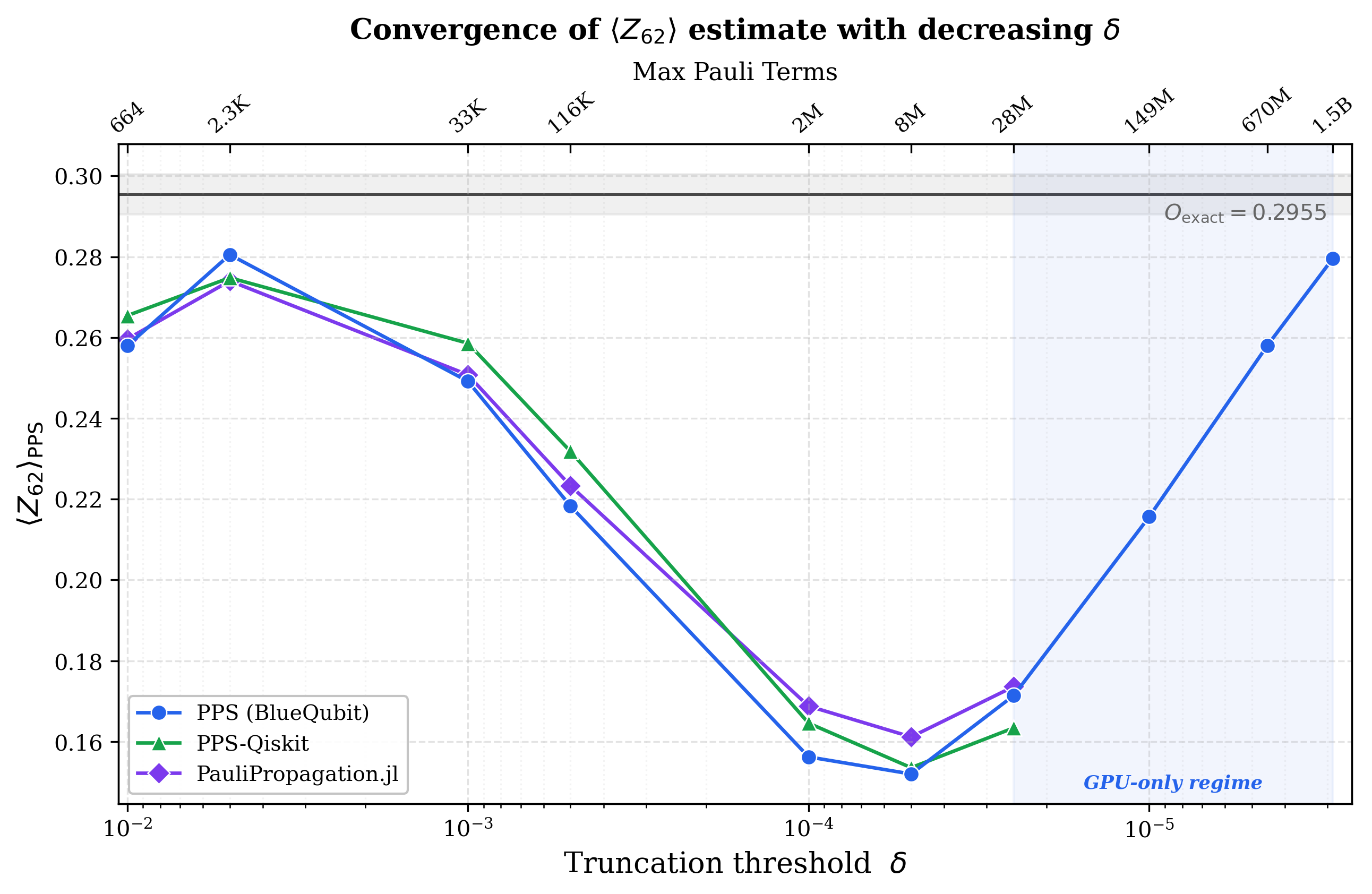}
    \caption{Raw PPS estimate $\langle Z_{62}\rangle_{\mathrm{PPS}}$ vs.\ $\delta$. All implementations agree where their $\delta$ ranges overlap, validating correctness. The estimate follows a non-monotonic trajectory: it rises slightly to ${\approx}\,0.280$ at $\delta = 5\times10^{-3}$, undershoots to a minimum of ${\approx}\,0.152$ near $\delta = 5 \times 10^{-5}$, then recovers monotonically through the GPU-only regime toward $O_{\mathrm{exact}} = 0.2955$~\cite{begusic2024converged}, reaching ${\approx}\,0.280$ at the finest threshold tested.}
    \label{fig:pps-convergence}
\end{figure}
 
\subsection{Runtime Comparison Across Backends}
\label{sec:pps-runtime}
 
\Cref{fig:pps-runtime} shows runtime as a function of $\delta$ for BlueQubit PPS-CPU, the two CPU-only SDKs Qiskit pauli-prop and PauliPropagation.jl, and BlueQubit PPS-GPU on H100, MI300X, and H200, with NVIDIA cuPauliProp overlaid at the three published thresholds.
 
\textbf{CPU-favored regime.} At $\delta \geq 5 \times 10^{-3}$ (fewer than 2,300 Pauli terms), CPU-based backends are faster: Qiskit pauli-prop completes in under 0.1\,s and PauliPropagation.jl in under 0.2\,s, while BlueQubit GPU incurs a fixed kernel-launch overhead of ${\sim}1$\,s across the three SKUs. 
BlueQubit CPU (${\sim}0.9$\,s) is also faster than GPU in this regime.
 
\textbf{Crossover and growing advantage.} The GPU crossover occurs near $\delta \approx 10^{-3}$. Beyond this point, GPU advantage grows rapidly: at $\delta = 10^{-4}$ (2.2M Paulis), BlueQubit GPU finishes in ${\sim}1.05$--$1.21$\,s across H100/MI300X/H200 (${\sim}210\times$--$310\times$ vs.\ BlueQubit CPU and Qiskit pauli-prop). At $\delta = 2.5 \times 10^{-5}$ (27.6M Paulis), the finest threshold reachable by all CPU implementations, the three GPUs complete in $3.17$\,s (H200), $3.23$\,s (MI300X), and $3.52$\,s (H100), versus 5,714\,s for BlueQubit CPU (${\sim}1{,}620\times$--$1{,}800\times$ speedup), 5,645\,s for Qiskit pauli-prop, and 55,430\,s (${\sim}15$\,h) for PauliPropagation.jl. No CPU implementation evaluated here is able to operate below $\delta = 10^{-5}$, but for two distinct reasons. The local Qiskit pauli-prop and PauliPropagation.jl baselines hit the 16\,GB physical-memory ceiling of the laptop described in \Cref{sec:setup}: working sets spill to the page file and runtimes inflate past any tractable schedule. BlueQubit CPU, by contrast, does not run out of memory. The platform's API currently rejects \texttt{pauli\_path\_truncation\_threshold} values smaller than $10^{-5}$ as out of supported range, so this is a software-imposed limit rather than a hardware one. Neither limit is fundamental to CPU-based PPS: large-memory CPU clusters with appropriate software support could in principle reach finer thresholds.

\textbf{Comparison against cuPauliProp.} NVIDIA's cuPauliProp~\cite{lowell2025cuquantum} is the only other GPU-accelerated Pauli propagation implementation with published figures on this circuit, reported as speedups over Qiskit pauli-prop of $56\times$, $108\times$, and $177\times$ at $\delta = 10^{-4}$, $5 \times 10^{-5}$, and $2.5 \times 10^{-5}$. Applying those factors to our measured Qiskit pauli-prop runtimes places cuPauliProp at $4.6$\,s, $9.4$\,s, and $31.9$\,s at the three thresholds, against ${\sim}1.05$--$1.21$\,s, ${\sim}1.41$--$1.58$\,s, and ${\sim}3.17$--$3.52$\,s for BlueQubit PPS-GPU. BlueQubit is therefore ${\sim}3.8\times$--$4.4\times$ faster at $\delta = 10^{-4}$, ${\sim}5.9\times$--$6.6\times$ at $5 \times 10^{-5}$, and ${\sim}9.1\times$--$10.1\times$ at $2.5 \times 10^{-5}$: the margin widens with $N_P$, the same trend seen against the CPU baselines. Two caveats apply. The cuPauliProp figures are inferred from published speedup ratios rather than measured by us, and the accelerator on which they were obtained is not stated in the source, so this is a comparison at matched $\delta$ against reported numbers, not a controlled same-hardware head-to-head. It does, however, establish that the GPU advantage reported here is not simply the generic benefit of moving PPS to a GPU.

\textbf{Hardware portability of PPS-GPU.} In the GPU-only regime the three accelerators track each other closely through $\delta = 4.5\times10^{-6}$ (${\sim}670$M Paulis; ${\sim}73$\,s on H200, ${\sim}90$\,s on H100, ${\sim}98$\,s on MI300X). Beyond that point the measured reach differs by SKU: H100 stops at $\delta = 4.5\times10^{-6}$ in this sweep, H200 continues to $\delta = 3.28\times10^{-6}$ (${\sim}1.21$B Paulis, ${\sim}143$\,s), and MI300X reaches $\delta = 2.89\times10^{-6}$ (${\sim}1.53$B Paulis, ${\sim}322$\,s). Absolute runtimes therefore depend on the accelerator, but the qualitative conclusion, that only GPU backends unlock the sub-$10^{-5}$ accuracy-recovery regime, is stable across NVIDIA and AMD hardware.
 
\textbf{Why the speedup grows.} The growing advantage follows from the parallel structure of PPS updates: each Pauli term's coefficient truncation, phase accumulation, and rescaling are independent operations that map directly to GPU SIMT execution. As $N_P$ grows, GPU utilization increases while per-term overhead is amortized, producing a speedup that increases monotonically with problem size. The log-log slope of the GPU runtime curves is markedly shallower than any CPU baseline, confirming a lower asymptotic scaling exponent.
 
\subsection{Accuracy and the Convergence Regime}
\label{sec:pps-accuracy}
 
\Cref{fig:pps-convergence,fig:pps-accuracy} reveal a tradeoff between runtime and accuracy that is invisible to any single-backend evaluation.
 
\textbf{Non-monotonic convergence.} The PPS estimate of $\langle Z_{62} \rangle$ does not converge monotonically toward the exact value $O_{\mathrm{exact}} = 0.2955$ as $\delta$ decreases. We adopt this reference value from the converged MIX tensor network simulations of Begu\v{s}i\'{c}, Gray, and Chan~\cite{begusic2024converged}, whose highest-accuracy results (effective wave-function and operator sandwich bond dimension ${>}\,16{,}000{,}000$) achieve absolute error ${<}\,0.01$; compatible values were independently obtained by Tindall et al.~\cite{tindall2024efficient}. Instead, the estimate is non-monotonic: starting at ${\approx}\,0.258$ at $\delta = 10^{-2}$, it rises briefly to ${\approx}\,0.280$ at $\delta = 5\times10^{-3}$, then undershoots to a minimum of ${\approx}\,0.152$ near $\delta = 5\times10^{-5}$, before recovering monotonically through the GPU-only regime to ${\approx}\,0.280$ at the finest threshold tested, $\delta = 2.89\times10^{-6}$.
 
This non-monotonic behavior is consistent with the convergence framework of Gharibyan et al.~\cite{gharibyan2025practical}, which identifies parameter regimes where reducing $\delta$ initially worsens PPS accuracy before eventually improving it. Our data provides direct empirical support: the trajectory produces a corresponding peak in accuracy error (\Cref{fig:pps-accuracy}), with $\epsilon$ first dipping to ${\approx}\,0.015$ at $\delta = 5\times10^{-3}$, rising to a peak of ${\approx}\,0.14$ near $\delta = 5\times10^{-5}$, then decreasing monotonically through the GPU-only regime to $\epsilon \approx 0.016$ at the finest threshold tested ($\delta = 2.89\times10^{-6}$).
 
\textbf{GPU unlocks the accuracy recovery regime.} The accuracy recovery from $\epsilon \approx 0.14$ back down to $\epsilon \approx 0.016$ at $\delta = 2.89\times10^{-6}$ occurs entirely in the regime that only BlueQubit GPU can reach within practical wall-clock time. This is the main finding of the cross-platform comparison: a practitioner limited to the CPU-based PPS backends evaluated here would observe accuracy degrading as $\delta$ decreases, with no on-platform path to finer thresholds, whether due to local-machine memory pressure or a vendor-imposed cap on the truncation parameter. As a representative GPU-only data point, the $\delta = 4.5\times10^{-6}$ computation completes in ${\sim}73$--$98$\,s across H200/H100/MI300X, making the convergence regime accessible for iterative experimentation; extrapolating CPU scaling curves, the same computation would require days of single-core compute.

\section{Discussion}
\label{sec:discussion}

\subsection{Cross-Platform Summary}
 
The benchmarks reveal a consistent pattern: BlueQubit GPU is the fastest zero-setup backend whenever problem size is sufficient to amortize GPU overhead: 1--2 orders of magnitude faster than Braket SV1 and Quantum Rings at 30+ qubits for state-vector, $2\text{--}3\times$ faster than CPU at $\chi = 256$ for MPS (with advantage growing as $\chi^{0.55}$), and up to ${\sim}1{,}700\times$ faster than BlueQubit CPU at $\delta = 2.5\times10^{-5}$ for PPS across H100, MI300X, and H200. The same protocol also identifies regimes where alternatives are competitive or superior: CPU backends outperform GPU for low-entanglement MPS circuits ($\chi \lesssim 128$) and coarse-threshold PPS ($\delta \geq 10^{-3}$). These crossover points, visible only through systematic comparison, are among the most practically useful outputs of this study. We stress that these rankings mix the two comparison types defined in \Cref{sec:setup}. The BlueQubit CPU-versus-GPU results isolate hardware on a fixed implementation, whereas the cross-vendor results also reflect implementation and provisioning differences and should be read as end-to-end platform comparisons.

\subsection{Architectural Interpretation of MPS Scaling}
 
The GPU exponent of $\chi^{1.49}$ reflects differential parallelization across constituent operations: matrix contractions map to high arithmetic-intensity GEMM kernels where GPU throughput scales favorably with matrix size, while batched SVD reduces the effective per-decomposition overhead. Published profiling of comparable implementations provides independent architectural support for this interpretation: Schieffer et al.~\cite{schieffer2025harnessingcudaqsmpstensor} put SVD at $\sim$70\% of GPU build time, TN-Sim~\cite{hoyt2026tnsim} reports similar dominance patterns, and cuTensorNet~\cite{bayraktar2023cuquantum} reports order-of-magnitude primitive speedups.

\subsection{Reproducibility and Standardization}
 
Vendor-reported numbers typically cover different circuits, parameter ranges, and metrics, making direct comparison impossible. By evaluating all backends on the same circuit families, parameter sweeps, and statistical protocol, and by releasing all code publicly, we aim to establish a reusable baseline. To let others recreate the environment, the repository also includes pinned dependency manifests for the local baselines, namely a Python \texttt{requirements.txt} with exact package versions for Qiskit pauli-prop and a Julia \texttt{Project.toml} and \texttt{Manifest.toml} for PauliPropagation.jl, together with the recorded SDK and API versions for the cloud backends and the hardware and OS specifications of \Cref{sec:setup}. The cloud-hosted backends depend on the provider's managed stack, which we identify by version where the platform exposes it and otherwise flag as a source of variability outside our control.

\subsection{Cost Considerations}

Cost matters to users of zero-setup platforms, and we do not fully quantify it here. Both AWS Braket SV1 and BlueQubit bill by compute time, so to first order the relative dollar cost tracks the relative runtimes reported throughout. The state-vector runtime gap at 30+ qubits (\Cref{fig:cross-platform-d60}) therefore implies a comparable cost gap, and the SV1 runtimes above $10^{7}$\,ms that led us to stop at 31 qubits correspond to proportionally high on-demand charges. We avoid a one-sided cost claim on purpose. A like-for-like cost-to-solution comparison would require normalizing pricing models, instance tiers, and free-tier allowances across platforms, which we leave to future work. We flag the omission explicitly because the decision to stop SV1 was driven by both runtime and cost, and a symmetric cost accounting for every backend, including BlueQubit, is not yet provided.

\subsection{Limitations}

The MPS $\chi$ range of 32--1536 does not yet reach the SVD-dominated regime; all BlueQubit results were obtained on a single managed platform, so the specific GPU architecture and software stack may influence the observed exponents; circuit topology affects MPS performance, as the QFT results demonstrate; and the non-monotonic PPS convergence observed on the kicked Ising circuit may not generalize to all circuits; Gharibyan et al.~\cite{gharibyan2025practical} identify families where PPS converges monotonically. Extending the benchmark to additional simulators (NVIDIA's CUDA-Q, qsim with MPS support) would strengthen the cross-platform conclusions.

\textbf{Hardware asymmetry and comparison scope.} The cross-vendor results combine implementation and hardware differences. BlueQubit's state-vector and MPS GPU backends run on A100 hardware; the PPS-GPU curves in \Cref{fig:pps-runtime} additionally report H100, MI300X, and H200 for the same engine; BlueQubit CPU backends use 2 to 20 cores; and the self-contained SDKs run on a 16\,GB commodity laptop. The CPU-to-GPU speedups are therefore upper bounds relative to well-provisioned, many-core CPU systems, and the cross-vendor rankings should be read as end-to-end platform comparisons rather than as measurements of algorithmic efficiency on its own. The BlueQubit CPU-versus-GPU comparisons, and the same-implementation MPS comparison in particular, are the cleanest infrastructure measurements in the study.

\textbf{Baseline selection.} AWS is represented only by the managed SV1 state-vector simulator. GPU-backed AWS workflows through Braket Hybrid Jobs~\cite{braket_jobs}, along with the DM1 and TN1 managed simulators, were not evaluated and could narrow the state-vector gap.

\textbf{Cost.} As discussed above, a symmetric cost-to-solution accounting across platforms is not provided.

\textbf{Workload coverage.} The suite omits hybrid variational workloads, such as variational quantum algorithms and quantum machine-learning circuits, whose alternating classical and quantum structure exercises CPU and GPU together, and it does not report circuit compilation or transpilation time. Adding these workloads and metrics, and integrating the suite with a community benchmarking frontend such as MQT Bench~\cite{quetschlich2023mqtbench}, are natural extensions.

 \subsection{Implications for Quantum Advantage}
 
The combination of MPS sub-quadratic scaling and PPS ${\sim}1{,}700\times$ speedup on GPU-accelerated zero-setup backends raises the classical simulation baseline against which quantum advantage must be demonstrated. For PPS, the error peak can now be traversed in minutes rather than days, which moves classical computations that were previously impractical into routine use. For MPS, the growing GPU advantage at high $\chi$ means that bond dimensions relevant to large-scale tensor-network simulations of utility-scale circuits ($\chi \sim 5000$, see e.g.~\cite{begusic2024converged}) may be reachable with projected runtimes of $\sim$12 hours on a zero-setup GPU backend, compared to $\sim$120 hours on CPU.

\section{Conclusion}
\label{sec:conclusion}
 
We have presented a systematic, cross-platform benchmarking study of zero-setup quantum circuit simulators across state-vector, MPS, and PPS methods. BlueQubit GPU is the fastest backend at scale: 1--2 orders of magnitude faster than Braket SV1 and Quantum Rings at 30+ qubits for state-vector, with sub-quadratic MPS scaling ($T \propto \chi^{1.49}$) producing a growing $\chi^{0.55}$ advantage, and up to ${\sim}1{,}700\times$ speedup on the 127-qubit kicked Ising PPS benchmark, where it is the only tested backend able to traverse the non-monotonic convergence regime of Gharibyan et al.~\cite{gharibyan2025practical}. A controlled QFT comparison shows that GPU kernel-launch overhead inverts this picture at low bond dimension ($\chi \lesssim 128$, up to $7.5\times$ slowdown), which makes circuit entanglement structure, not qubit count, the axis that should drive backend selection. Beyond identifying the fastest backend in each regime, the primary contribution is methodological: a standardized protocol that exposes crossover points invisible to single-platform evaluation.

\section*{Acknowledgments}

We thank Tigran Mamikonyan for assistance with the MPS simulators, including
resolving backend issues and providing the peaked circuits; Greg Vardanyan for
support with state-vector and MPS simulator issues; and David Sargsyan for
general support and for connecting us with the appropriate experts on more
specialized questions.

\textbf{Conflict of Interest.} All authors are employed by or affiliated with
BlueQubit, one of the simulation platforms evaluated in this work. To mitigate
potential bias, all BlueQubit backends were evaluated under the identical
circuit families, parameter sweeps, and statistical protocol applied to every
competing platform. All benchmarking code and raw results are publicly
available for independent verification.\footnote{\repo}

\textbf{AI Assistance.} Large language model assistance (Claude, Anthropic) was
used for manuscript editing and structural revision. All scientific content,
experimental data, and analysis are the authors' own work.

\bibliographystyle{unsrt}
\bibliography{references}
\end{document}